\documentclass[%
reprint,
superscriptaddress,
preprintnumbers,
amsmath,amssymb,
aps,
prd,
]{revtex4-1}

\usepackage[utf8]{inputenc}

\usepackage{mathtools}
\usepackage{amsfonts}
\usepackage{mathrsfs}
\usepackage{bm}
\usepackage{bbold}
\usepackage{slashed}
\usepackage{dsfont}
\usepackage{float}
\usepackage{dcolumn}
\usepackage{amssymb,amsmath}

\usepackage{graphicx}
\usepackage{subcaption}
\usepackage{color}
\usepackage{array}

\usepackage{placeins}
\usepackage{booktabs}
\usepackage{caption}

\usepackage{xspace}
\usepackage{hyperref}
\usepackage[nameinlink]{cleveref}
\usepackage{bookmark}
\usepackage{units}

\usepackage{ulem}

\usepackage{booktabs}
\usepackage{multirow}
\newcolumntype{C}{>{$}c<{$}}
\AtBeginDocument{
	\heavyrulewidth=.08em
	\lightrulewidth=.05em
	\cmidrulewidth=.03em
	\belowrulesep=.65ex
	\belowbottomsep=0pt
	\aboverulesep=.4ex
	\abovetopsep=0pt
	\cmidrulesep=\doublerulesep
	\cmidrulekern=.5em
	\defaultaddspace=.5em
}

\usepackage{xifthen}
\usepackage{xcolor}

\newcommand{\gettitle}{Chemical freeze-out parameters via functional renormalization group approach}

\hypersetup{
	colorlinks,
	linkcolor={red!75!black},
	citecolor={blue!75!black},
	urlcolor={blue!75!black},
	pdftitle={\gettitle},
	pdfauthor={},
	pdfkeywords={}
	{}
	bookmarksopen=true,
	bookmarksopenlevel=2,
	bookmarksnumbered=true
}

\begin{document}

\title{Chemical freeze-out parameters via functional renormalization group approach}
	
 \author{Jun-xiang Shao}\email{junxiangshao@pku.edu.cn}
 \affiliation{Department of Physics and State Key Laboratory of Nuclear Physics and Technology, Peking University, Beijing 100871, China}
 \affiliation{Collaborative Innovation Center of Quantum Matter, Beijing 100871, China}

 \author{Wei-jie Fu}\email{wjfu@dlut.edu.cn}
 \affiliation{School of Physics, Dalian University of Technology, Dalian, 116024, China}

 \author{Yu-xin Liu} \email[Corresponding author: ]{yxliu@pku.edu.cn}
 \affiliation{Department of Physics and State Key Laboratory of	Nuclear Physics and Technology, Peking University, Beijing 100871, China}
 \affiliation{Collaborative Innovation Center of Quantum Matter, Beijing 100871, China}
 \affiliation{Center for High Energy Physics, Peking University, Beijing 100871, China}

\begin{abstract}

We study the freeze-out parameters in a QCD-assisted effective theory that accurately captures the quantum and in-medium effects of QCD at low energies. Functional renormalization group approach is implemented in our work to incorporate the non-perturbative quantum, thermal and density fluctuations. By analyzing the calculated baryon number susceptibility ratios $\chi_{2}^{B}/\chi_{1}^{B}$ and $\chi_{3}^{B}/\chi_{2}^{B}$, we determine the chemical freeze-out temperatures and baryon chemical potentials in cases of hard thermal or dense loop improved $\mu$-dependent glue potential and $\mu$-independent glue potential. We calculate the ${\chi_{4}^{B}}/{\chi_{2}^{B}}\, (\kappa \sigma^{2})$ and ${\chi_{6}^{B}}/{\chi_{2}^{B}}$ along the freeze-out line for both cases. It's found that $\kappa \sigma^{2}$ exhibits a nonmonotonic behavior in low collision energy region and approach to one for lower collision energy. ${\chi_{6}^{B}}/{\chi_{2}^{B}}$ shows a similar complicated behavior in our calculation.

\end{abstract}

\maketitle

\section{introduction}
\label{sec:Intro}

The study of QCD phase transitions is a very active field of research. The transitions include both the chiral phase transition and the deconfinement phase transition. The chiral phase transition is signaled by the dynamical chiral symmetry breaking and responsible for the mass of visible matter in the universe, while the deconfinement phase transition is signaled by the center symmetry breaking and related to hadron formulation. A thorough understanding of QCD phase transitions may shed light on the process and the nature of the early universe evolution~\cite{Schwarz:2003du}.
The first principle lattice QCD simulations indicate that the phase transition is a smooth crossover in the low baryon chemical potential~\cite{HotQCD:2018pds,HotQCD:2019xnw}. Although its predictive capability is hampered due to the notorious sign problem region~\cite{Muroya:2003qs,Splittorff:2007ck} when the finite baryon chemical potential is considered, other first principle methods as well as low energy effective theories, such as the functional renormalization group (FRG)~\cite{Fu:2022gou,Fu:2021oaw,Fu:2019hdw,Fu:2016tey,Chen:2021iuo,Pawlowski:2014zaa,Rennecke:2016tkm,Fu:2018qsk}  and Dyson-Schwinger equations (DSE)~\cite{Isserstedt:2019pgx,Fischer:2011mz,Xin:2014ela,Gao:2020qsj} have come to play a complementary role, and some evidence has been found that there might be a critical end point (CEP) in the temperature $T $--baryon chemical potential $\mu_{B}$ plane. The existence of CEP still needs confirmation in experiments, and its exact position if it exists, have become one of the most significant topics in both theories and experiments~\cite{Davis:2020fcy,Odyniec_2010,STAR:2013gus,Luo:2015ewa}.

In experiments, however, one can not measure the phase transitions directly but only the hadron states after hadronization, to wit, the chemical freeze-out state which is defined as the set of particle states ceasing the inelastic collisions between hadrons. For different collision energies one obtains different chemical freeze-out states corresponding to different chemical freeze-out points in the $T $--$\mu_{B} $ plane, and thus one observes a chemical freeze-out line connecting different freeze-out points in the plane.
It has been shown that a nonmonotonic behavior of conserved charge fluctuations can arise as the chemical freeze-out line approaches to the CEP~\cite{Stephanov:1999zu,Hatta:2003wn,Stephanov:2011pb}. The Beam Energy Scan (BES) program at the Relativistic Heavy Ion Collider (RHIC), the Facility for Antiproton and Ion Research (FAIR) in Darmstadt, the Nuclotron-based Ion Collider Facility (NICA) in Dubna and the high-intensity heavy-ion accerator facility (HIAF) in Huizhou all take the search of the CEP as one of their most important scientific objectives~\cite{Odyniec_2010,Davis:2020fcy,Zhou:2022pxl}, and some important results have been found by the RHIC collaboration~\cite{Luo:2015ewa,STAR:2020tga,STAR:2022vlo,STAR:2022hbp,STAR:2019ans,STAR:2017tfy}.
In theoretical aspects, the freeze-out conditions have been studied in Dyson-Schwinger equations (DSE) approach~\cite{Isserstedt:2019pgx,Lu:2021ium},  functional renormalization group (FRG) approach~\cite{Fu:2015amv,Fu:2021oaw}, lattice QCD simulations~\cite{Borsanyi:2014ewa,Bazavov:2012vg,Gavai:2010zn}, statistical hadronization models~\cite{Andronic:2005yp,Andronic:2017pug,Becattini:2005xt,Alba:2014eba,Karsch:2010ck} and other models~\cite{Chen:2015dra,Fukushima:2010is,Li:2018ygx}. 
However, the position of the chemical freeze-out line, the existence and the location (if exsists) of the CEP, the conservation charge fluctuations, and some other problems are still unclear and further investigations are necessary, especially when the high baryon chemical potential is involved.
We then take the FRG approach combining with a QCD-assisted effective field theory model to study the position of the chemical freeze-out line, the higher order baryon number fluctuations and related issues in finite temperature and baryon chemical potential system in this paper.

The QCD-assisted low-energy effective theory naturally emerges from QCD by integrating out the high energy degrees of freedom at the low energy scale. We choose the Polyakov-quark-meson (PQM) model as an effective realization of QCD in this work, 
and make use of the FRG approach to deal with the effect of the fluctuations of the fields and other nonperturbative properties of the QCD system. 
The PQM model has extended the traditional quark-meson model that describes chiral dynamics of QCD quite well to include the description of the deconfinement aspects of QCD phase structures~\cite{Fukushima:2003fw,Fukushima:2017csk,Schaefer:2007pw,Herbst:2010rf}, and it is also well-suited for the study of thermodynamical properties and baryon number fluctuations in QCD system.
It has also been known that the low-energy effective models can be related to full QCD systematically within the FRG approach~\cite{Braun:2011pp,Pawlowski:2005xe,Berges:2000ew}, see also the discussion in Sec.~\ref{sec:PQM}.

The one vital component of the PQM model is the construction of the glue potential of full QCD, which encodes the glue dynamics in the presence of matter fields and is represented effectively by a Polyakov-loop potential. The potential is fixed by fitting the lattice QCD data of the pure Yang-Mills system without quarks~\cite{Lo:2013hla,Roessner:2006xn,Fukushima:2017csk}, and thus the coupling of the matter sector to the gauge sector is lost. There have been many attempts on recovering this unquenched effect and some significant progress has been made at least for small $\mu_{B}$~\cite{Schaefer:2007pw,Haas:2013qwp,Herbst:2013ail,Herbst:2013ufa,Stiele:2016cfs}.
In this work we simply take two kinds of Polyakov-loop potentials. 
One is $\mu$-dependent through $\mu$-dependent pseudocritical deconfinement phase transition temperature $T_{0} $, 
which is based on hard thermal or dense loop consideration and first put forward in Ref.~\cite{Schaefer:2007pw} for a better treatment of the phase structures in the finite density situations, see Eqs.~(\ref{eq:T0mu}) and (\ref{eq:bmu}) for the specific form used in this work.
The $\hat{\gamma}$ in Eq.~(\ref{eq:bmu}) has been chosen to be one in Refs.~\cite{Fu:2018swz,Fu:2018qsk}, 
and the calculated results have found a good agreement with the results of lattice QCD and the HRG model. 
The other is $\mu$-independent simply by choosing $T_0 $ as a constant with no hard thermal or dense loop improvements incorporated. The results based on these two different potentials are compared and discussed in the main text. The influence of other different values of $\hat{\gamma}$ is also explored and discussed in Appendix~\ref{app:influ}.

Compared with previous works on QCD system of two flavors~\cite{Friman:2011pf,Almasi:2017bhq,Fu:2015amv,Fu:2021oaw}, 
the inclusion of the strange quark brings more mesons into the system so that the system is closer to the real one.
To include non-perturbative quantum, thermal and density fluctuations, the FRG approach is implemented at finite temperature and density.
As a nonperturbative continuum field approach, FRG has been successfully applied in first principle QCD and model calculations~\cite{Braun:2014ata,Cyrol:2017qkl,Mitter:2014wpa,Mitter:2013fxa,Fu:2019hdw,Fu:2021oaw}. It shows a powerful performance on non-perturbative problems.
  	
This paper is organized as follows. After this introduction, we briefly describe the PQM model from the FRG perspective in Sec.~\ref{sec:PQM}. 
In Sec.~\ref{sec:eff} we describe our theoretical framework and present our numerical setup. 
In Sec.~\ref{sec:results} we give our numerical results on low-order susceptibility ratios, freeze-out parameters and high-order susceptibility ratios. We put a discussion of the influence of other different values of $\hat{\gamma}$ in Appendix~\ref{app:influ}.
In Sec.~\ref{sec:summary} we give a brief summary of our obtained results.

\section{Functional renormalization group approach to the PQM model}
\label{sec:PQM}

\begin{figure}[htb]
\includegraphics[width=0.47\textwidth]{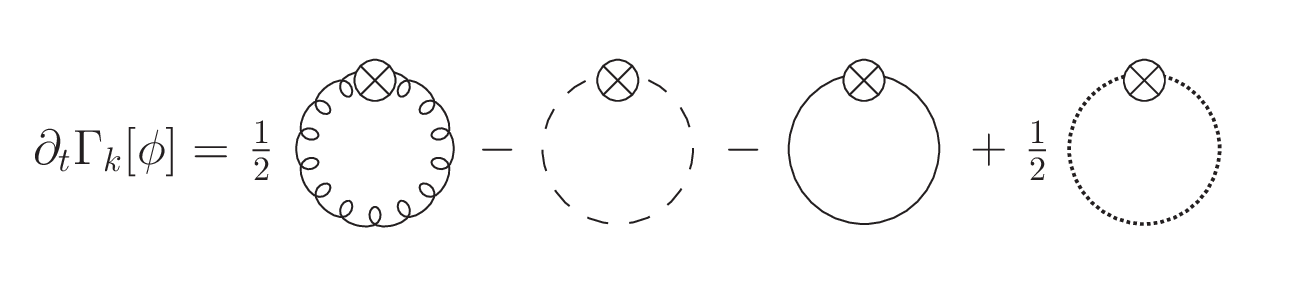}
\caption{Partially hadronized version of the FRG flow for the effective action in QCD. The loops denote the gluon, ghost, quark and meson contributions, respectively. The crosses mark the regulator term. }
\label{fig:QCDflow}
\end{figure}

FRG is a functional continuum field approach of QCD. 
A pictorial representation of the flow of QCD is shown in Fig.~\ref{fig:QCDflow}, 
in which the first three loops in the right hand side denote the gluon, ghost and quark contributions respectively, 
while the fourth is the mesonic loop introduced via the dynamical hadronization~\cite{Berges:2000ew,Glazek:2008mj}.
%
%

In the QM model, only the contributions from the last two loops in Fig.~\ref{fig:QCDflow} are taken into account, accounting only for the chiral aspect of QCD phase transitions without the deconfinement. 
As a better imitation of full QCD, the PQM model keeps all these four loops but implements an effective glue potential to represent the contributions from the first two loops.

%
\begin{figure}[htb]
	\includegraphics[width=0.3\textwidth]{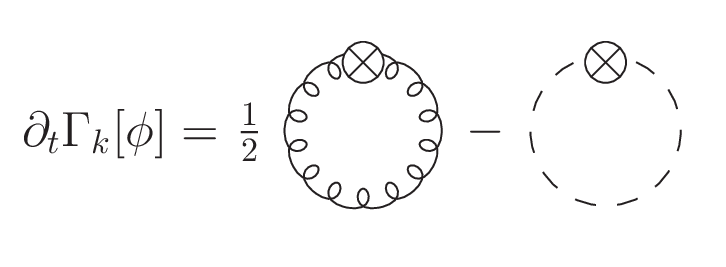}
	\caption{Flow equation of the effective action in the pure Yang-Mills theory.}
	\label{fig:YMflow}
\end{figure}	
%

The glue potentials are commonly phenomenologically constructed on the basis of considering the fundamental symmetries and only finite order terms of polynomials satisfying the symmetries are kept in their ansatz for simplicity. 
The parameters in the potentials are determined by fitting the lattice QCD results for the pressure, the entropy density, the energy density and the evolution of the expectation value of the Polyakov-loop with temperature in the pure gauge theory, see Refs.~\cite{Roessner:2006xn,Fukushima:2017csk,Lo:2013hla,Xin:2014dia,Scavenius:2002ru} for their specific forms and further details.
Note that the construction is done in the pure gauge theory, which corresponds to Fig.~\ref{fig:YMflow} in the formulation of FRG.
	
In Fig.~\ref{fig:QCDflow} the gluon propagator can receive a contribution from quarks by quark polarization. 
However, the contribution is missing in Fig.~\ref{fig:YMflow} due to the lack of quark freedoms in the pure gauge theory. 
The simplification to the contributions of the first two loops in Fig.~\ref{fig:QCDflow} by the effective action in Fig.~\ref{fig:YMflow} means the neglect of the backreaction of the matter sector to the gauge sector. 
Moreover, the above arguments also show that it is a formally additive structure for the different contributions, and thus allows us to systematically improve the low-energy effective models towards full QCD within the FRG approach.

One important effect of the inclusion of the dynamical quarks in full QCD is the change of the scale $\Lambda_{\rm QCD} $, to which the transition temperature $T_{0} $ of the Polyakov-loop potential is related. 
Based on the hard thermal or dense loop resumation results, a relation is proposed in Ref.~\cite{Schaefer:2007pw} to estimate the flavor and the quark chemical potential dependence of $T_0(N_f,\mu) $. 
This accounts partially for the backreaction effect. Furthermore, the $\mu $ dependence of $T_0 $  is modified to account for the Silver-Blaze property of QCD in \cite{Herbst:2013ail}. Another important observation comes from Ref.~\cite{Haas:2013qwp} that, although the QCD glue potential is different from the Yang-Mills potential their shapes are similar, and  the difference between them can be significantly decreased by a rescale of the  temperature, see Eq.~(\ref{eq:rescalet}) for the specific form. Based on this observation, a better agreement with lattice QCD results is achieved~\cite{Haas:2013qwp,Herbst:2013ufa}.

\section{2+1 flavor low energy effective model}
\label{sec:eff}
\subsection{Theoretical framework}
\label{subsec:frame}

The PQM model is an effective realization of QCD system~\cite{Schaefer:2007pw,Skokov:2010wb,Fukushima:2017csk}. The Lagrangian of the 2+1 flavor PQM in the Euclidean space is given as
\begin{align}\label{Lagrangian}
\mathcal{L} =& \bar{q}[\gamma_\mu\partial_\mu - \gamma_0(\hat\mu + igA_0)]q + h\bar{q}\Sigma_5q \nonumber\\
&+ {\rm Tr}(\bar{D}_\mu\Sigma\cdot\bar{D}_\mu\Sigma^\dagger) + \tilde{U}(\Sigma,\Sigma^{\dagger}) + V_{\rm glue}(L,\bar{L}) \, ,
\end{align}
where $q $ is the quark field with three flavors $(u,d,s) $ and three colors. $\hat{\mu} $ is the matrix form of the quark chemical potential in the flavor space, i.e. $\hat{\mu}={\rm diag}(\mu_u,\mu_d,\mu_s) $. The quark potentials are related to the baryon, isospin and strangeness chemical potentials as follows
\begin{align}
	\renewcommand\arraystretch{1.3}
\begin{pmatrix} \mu_u \\ \mu_d \\ \mu_s \end{pmatrix} =
\begin{pmatrix} \frac{1}{3}\mu_B+\frac{1}{2}\mu_I \\ \frac{1}{3}\mu_B-\frac{1}{2}\mu_I \\ \frac{1}{3}\mu_B-\mu_S \end{pmatrix} \,.
	\renewcommand\arraystretch{1.0}
\end{align}
 For the time being, we use $\mu_{I} = 0 $ and do not consider the strangeness neutrality requirement in experiments, to wit, $\mu_{S} =0 $ is used. It is found that the constraint of strangeness neutrality  begins to make a sizable difference only when $\mu_{B}$ is significantly large~\cite{Fu:2018qsk}.
	
Meson fields are combined into the matrix forms which read
\begin{align}
	\Sigma =& \sum_{a=0}^8\ (\sigma_a + i\pi_a)T^a \,, \nonumber\\
	\Sigma_{5}=& \sum_{a=0}^8\ (\sigma_a + i\gamma_5\pi_a)T^a \,,
\end{align}
where $\sigma_{a}$ and $\pi_{a}$ are the scalar and pseudo-scalar meson nonets, $T^{a}$ is the ${\rm SU(3)}$ generator in the flavor space, 
and $T^{0}= \frac{1}{\sqrt{6}} I_{3 \times 3}^{} $. The covariant derivative of meson fields is defined as
\begin{align}
\bar{D}_\mu\Sigma = \partial_\mu\Sigma + \delta_{\mu 0}[\hat{\mu},\Sigma] \,.
\end{align}
The meson potential $\tilde{U}(\Sigma,\Sigma^{\dagger})$ reads
\begin{align}
\tilde{U}(\Sigma,\Sigma^{\dagger}) = U(\rho_1,\rho_2) - j_l\sigma_L - j_s\sigma_S - c_A\xi \, ,
\end{align}
where $j_l \sigma_L $ and $j_s \sigma_S $ are explicit symmetry breaking terms, which reduce the $\rm SU_V(3) $ symmetry to the $\rm SU_V(2) $. $\sigma_L $ and $\sigma_S $ are related to meson fields via the chiral rotation
\begin{align}
	\begin{pmatrix}
		\sigma_L\\
		\sigma_S
	\end{pmatrix}
	=\frac{1}{\sqrt{3}}
	\begin{pmatrix}
		\sqrt{2} & 1\\
		1 & -\sqrt{2}
	\end{pmatrix}
	\begin{pmatrix}
		\sigma_0 \\
		\sigma_8
	\end{pmatrix}  \,.
\end{align}
$U(\rho_1,\rho_2) $ is symmetric under the $\rm U_V(3)\times U_A(3) $ transformations with two chiral invariants $\rho_1 $ and $\rho_2 $,
\begin{align}
\rho_{1}^{} =& {\rm Tr}\big [ \Sigma\cdot\Sigma^\dagger \big ] \,, \nonumber\\
\rho_{2}^{} =& {\rm Tr}\big [ (\Sigma\cdot\Sigma^\dagger - \frac{1}{3}\rho_1)^2 \big ] \, .
\end{align}
$\xi $ is the Kobayashi--Maskawa--'t Hooft determinant which originates from the $\rm U_A(1) $ anomaly effect~\cite{Kobayashi:1970ji,tHooft:1976rip} and reads
\begin{align}
\xi={\rm det}(\Sigma)+{\rm det}(\Sigma^\dagger)\,.
\end{align}

The temporal gluon background field $A_0 $ in Eq.~(\ref{Lagrangian}) can be formulated into the Polyakov loops, to wit,
\begin{align}
	L(\bm{x}) =	\frac{1}{N_c}  {\rm Tr}\,\mathcal{P}(\bm{x}) \,, \quad  \bar L (\bm{x})=\frac{1}{N_c}  {\rm Tr}\,{{\mathcal{P}}^{\dagger}(\bm{x})}
\end{align}
with
\begin{align}
	\mathcal{P}(\bm x) = \mathcal{P}\exp\Big(ig\int_0^{\beta}d\tau A_0(\bm{x},\tau)\Big) \,,
\end{align}
where $\mathcal{P} $ is the path ordering operator.

Once the Lagrangian is given, there are several methods to calculate the effective action of the system. Mean-field approximation is one of the widely used method, which treats meson fields as background fields and neglect their fluctuations. However, it is shown that mesonic fluctuations are important for physics at low temperature~\cite{Skokov:2010wb}. By contrast, the FRG approach can incorporate the fluctuations of mesons as well as quarks. It implements the procedure by introducing the regulator term $R_k(p) $, which suppresses the fluctuations for momenta $p<k$, and leaves the fluctuations unchanged for $p>k $. So when $k $ is near the cutoff $\Lambda $, no fluctuations are included, i.e. $\Gamma_{k=\Lambda}=S_{\rm bare} $. When $k $ approaches to zero, all fluctuations are included and we obtain the full effective action $\Gamma_{k=0}=\Gamma_{\rm Full} $. The ideas are formulated in the Wetterich equation~\cite{Wetterich:2001kra}, which reads
\begin{align}\label{WetterichEq}
\partial_{t}\Gamma_{k} =\frac{1}{2}{\rm Tr}\big(G^{\phi\phi}_{k}\partial_{t} R^{\phi}_{k}\big) - {\rm Tr}\big(G^{q\bar q}_{k}\partial_{t} R^{q}_{k}\big) \,,
\end{align}
where $t=\ln(k/\Lambda) $, $\phi $ represents the mesonic degrees of freedom; $G^{\phi\phi}_{k} $ and $G^{q\bar q}_{k} $ are the meson and quark propagators, respectively; $R^{\phi}_{k} $ and $R^{q}_{k} $ are the regulators for the meson and quark fields.

Similar with the DSE approach, FRG equations are infinitely coupled, so that truncations have to be taken to solve them. 
We truncate $\Gamma_k $ as follows and make use of the local potential approximation (LPA),
\begin{align}\label{TruncatedGammak}
\Gamma_k =& \int_x \bigg\{ \bar{q}[\gamma_\mu\partial_\mu - \gamma_0(\hat\mu + igA_0)]q \nonumber\\
&+ h\bar{q}\Sigma_5q + {\rm Tr}(\bar{D}_\mu\Sigma\cdot\bar{D}_\mu\Sigma^\dagger) +  U_k(\rho_1,\rho_2) \nonumber\\
&- j_l\sigma_L - j_s\sigma_S  - c_A\xi + V_{\rm glue}(L,\bar{L}) \bigg\} \,,
\end{align}
where only $U_{k}(\rho_{1},\rho_{2}) $ flows with the RG scale $k $. The wave function renormalizations of the fields and the Yukawa coupling do not run with the flow in this approximation.

Substituting Eq.~(\ref{TruncatedGammak}) into Eq.~(\ref{WetterichEq}) and implementing the optimized regulators in Ref.~\cite{Litim:2001up}, 
one obtains an analytic flow equation for $U_k(\rho_1,\rho_2) $~\cite{Wen:2018nkn,Fu:2018qsk},
\begin{align} \label{eq:uflow}
	&\partial_{t} U_k(\rho_1,\rho_2) 																	 	\nonumber\\
	& =\frac{k^4}{4\pi^2} \bigg\{ 3 l_0^{(B)}(\bar m_{a_0,k}^2,T,0) + 4l_0^{(B)}(\bar m_{\kappa,k}^2,T,0)	 \nonumber\\
	& + l_0^{(B)}(\bar m_{\sigma,k}^2,T,0) + l_0^{(B)}(\bar m_{f_0,k}^2,T,0) + 3 l_0^{(B)}(\bar m_{\pi,k}^2,T,0)    \nonumber\\
	& + 4 l_0^{(B)}(\bar m_{K,k}^2,T,0) + l_0^{(B)}(\bar m_{\eta,k}^2,T,0) + l_0^{(B)}(\bar m_{\eta^\prime,k}^2,T,0)  \nonumber\\
	& - 4N_c \Big[ 2 l_0^{(F)}(\bar m_{l,k},T,\frac{1}{3} \mu_B) + l_0^{(F)}(\bar m_{s,k},T,\frac{1}{3} \mu_B) \Big] \bigg\} \,,
\end{align}
with $\bar m_{i, k}^{2}\equiv m_{i, k}^{2}/k^2$. The quark masses are given by
\begin{align} \label{eq:qmass}
	m_l = \frac{h}{2} \sigma_L \,, \quad m_s = \frac{h}{\sqrt{2}}\sigma_S \,,
\end{align}
while the meson masses are obtained through diagonalizing the Hessian matrix $H_{ij} $,
\begin{align} \label{eq:hessian}
	H_{ij}=\frac{\partial^{2}\tilde{U}(\Sigma,\Sigma^\dagger)}{\partial\phi_i \partial\phi_j} \,,
\end{align}
with $\phi_i=(\sigma_1,\sigma_2,...,\sigma_8,\pi_1,\pi_2,...,\pi_8)$. The explicit expressions of meson masses can be found in Refs.~\cite{Kamikado:2014bua,Mitter:2013fxa}, for example.
The threshold functions $l_0^{(B/F)} $ in Eq.~(\ref{eq:uflow}) are collected in Appendix~\ref{app:eqs}.

The Polyakov-loop potential $V_{\rm glue} $ needs to be specified for the equations above. 
We take the parametrization in Ref.~\cite{Lo:2013hla}
\begin{align}\label{eq:Vglue}
\frac{V_{\rm glue}(L,\bar L)}{T^4} &= -\frac{a(T)}{2} \bar L L +  b(T)\ln M_H(L,\bar{L}) \nonumber\\
	&\quad + \frac{c(T)}{2} (L^3+\bar L^3) + d(T) (\bar{L} L)^2 \,,
\end{align}
which has the advantage of reproducing the Ployakov-loop susceptibilities as well as the usual thermal quantities of ${\rm SU(3)} $ Yang-Mills theory obtained in lattice QCD simulations. 
The specific form and coefficients of Eq.~(\ref{eq:Vglue}) are collected in Appendix~\ref{app:eqs}, see Eq.~(\ref{eq:haar}) to (\ref{eq:tYM}) and Table \ref{table:coeffs}. The $t$ in Eq.~(\ref{eq:haar}) to (\ref{eq:tYM}) is the reduced temperature in the pure gauge theory, one needs to rescale it to account for the unquenched effect when quarks are included~\cite{Haas:2013qwp}
\begin{align} \label{eq:rescalet}
	t=\frac{T-T_{\rm YM}}{T_{\rm YM}} \Longrightarrow \alpha\frac{T-T_0}{T_0} \,.
\end{align}
The values of the $\alpha$ and $T_{0}$ are explored and discussed in Ref.~\cite{Haas:2013qwp}. Since they have some dependence on the number of quark flavors and the parameterization of the Polyakov loop potential, we treat them as free parameters and determine them by fitting the pressure, trace anomaly with the lattice results. We choose the same values as those in Ref.~\cite{Fu:2018qsk}, which are also presented in Table~\ref{table:parameters}.

As mentioned in Sec.~\ref{sec:PQM}, $\mu $-dependent $T_0 $ is put forward to account further for the unquenched effect of the Polyakov-loop potential. Based on the identification of $\Lambda_{\rm QCD} $ in the one-loop beta function of QCD at large density with the modification of the critical temperature, a modification of $T_{0} $ is suggested as Ref.~\cite{Schaefer:2007pw}
\begin{align} \label{eq:T0mu}
	T_0(N_f,\mu) = T_\tau e^{-1/(\alpha_0 b_\mu)} \,,
\end{align}
where the renormalization scale is given by $T_\tau=1.77 \,$GeV with the strong coupling $\alpha_{0}=0.304$. 
The $\mu $ dependence is encoded in $b_{\mu} $ as~\cite{Fu:2018qsk},
\begin{align} \label{eq:bmu}
	b_\mu = b_0 - \frac{16}{\pi} \bigg[ 2\frac{\mu^2}{(\hat\gamma T_\tau)^2} \Delta n_l + \frac{\mu^2}{(\hat\gamma T_\tau)^2} \Delta n_s \bigg] \,.
\end{align}
In Ref.~\cite{Schaefer:2007pw}, $b_0$ is chosen as one-loop QCD beta function coefficient with $b_{0} = (11 N_{c} -2 N_{f})/(6\pi)$, 
and the second part in Eq.~(\ref{eq:bmu}) with $\hat{\gamma}=1$ and $\Delta n_{l/s}=1$ is constructed such that the deconfinement phase transition and the chiral phase transition coincide for finite density at mean field level. 
In this work, we treat $b_{0}$ as free parameters in the same way as $T_{0}$, and choose $\Delta n_{l/s}$ as follows in order to account for the Silver-Blaze property of QCD~\cite{Herbst:2013ail,Fu:2018qsk},
\begin{align} \label{eq:delnls}
\Delta n_{l/s} =& \frac{1}{e^{3(m_{l/s}-\mu)/T}+1} + \frac{1}{e^{3(m_{l/s}+\mu)/T}+1} \nonumber\\
&- \frac{2}{e^{3 m_{l/s}/T}+1} \,,
\end{align}
where $m_{l/s}$ is the vacuum mass of the light/strange quark, and $\Delta n_{l/s} $ reduces to the step function $\theta(\mu-m_{l/s}) $ at vanishing temperature.

The $\hat{\gamma} $ in Eq.~(\ref{eq:bmu}) controls the curvature of the deconfinement phase transition. 
For $\hat{\gamma}=\infty $, $b_\mu $ in Eq.~(\ref{eq:bmu}) is a constant which means no dependence on $\mu$, 
and the Polyakov-loop potential becomes $\mu $-independent.
For finite values of $\hat{\gamma} $ the Polyakov-loop potential will develop a direct dependence on $\mu $, which can be viewed as the hard thermal or dense loop improved Polyakov-loop potential when the finite density is involved. In Ref.~\cite{Fu:2018qsk} $\hat{\gamma} $ is chosen to be one and the obtained results show a good agreement with the lattice results for the equation of state. 
With the same setup, the baryon-strangeness correlation is calculated in Ref.~\cite{Fu:2018swz}, 
and the obtained results are in agreement with those of the HRG model. 
Thus in this work we show and compare the results calculated with $\hat{\gamma}=\infty $ and $\hat{\gamma}=1 $, and denote the latter case as the $\mu$-dependent case. In Appendix~\ref{app:influ} we also discuss the influence of other different values of $\hat{\gamma}$.

\subsection{Numerical setup}
\label{sec:numerical}

There are several methods to solve the flow equation in Eq.~(\ref{eq:uflow}). 
The most commonly used method is the Taylor expansion around the scale dependent minimum of the effective potential (the so-called running Taylor method), which is numerically labour-saving but has some drawbacks. 
Its convergence property is suspicious~\cite{Pawlowski:2014zaa} and it may not allow for the value of the $\sigma $ meson mass below $500\,$MeV due to numerical instabilities~\cite{Wen:2018nkn}. See also Ref.~\cite{Yin:2019ebz} for more discussions about the properties of convergence for this method. Another used approach is the grid method, which discretizes the fields ($\rho_{1},\rho_{2} $) on the multi-dimensional grid and replace the derivatives of $U_{k} $ with respect to the fields with appropriate numerical diffenences, and then the flow equation can be transformed into a set of coupled ordinary differential equations~\cite{Mitter:2013fxa,Resch:2017vjs}.
The grid method can capture the global properities of the effective potential but requires more numerical efforts. A fixed-point Taylor expansion has been developed in Refs.~\cite{Pawlowski:2014zaa,Rennecke:2016tkm}. In contradistinction to the running Taylor method whose convergence property is hampered by the linear feedback from the high-order expansion coefficients to the lower ones~\cite{Pawlowski:2014zaa,Yin:2019ebz}, the potential is expanded around a scale-independent point and a better convergence is obtained. More specifically, we expand the $U_k $ in Eq.~(\ref{eq:uflow}) as
\begin{align} \label{eq:uexpand}
U_k(\rho_1,\rho_2) =\sum_{n+2m=0}^N\frac{\omega_{nm,k}(T,\mu)}{n!m!}(\rho_1-\kappa_1)^n(\rho_2-\kappa_2)^m \,,
\end{align}
where expansion point $\kappa_1 $ and $\kappa_2 $ are scale-independent and should be the IR minimum of the effective potential. Setting $N=5 $ has been shown to be sufficient for numerical convergence~\cite{Rennecke:2016tkm,Fu:2018qsk,Fu:2018swz}. The equations for the coefficients are
\begin{align}  \label{eq:coeflow}
\partial_k\omega_{nm,k}(T,\mu) = \frac{\partial^{n+m} \partial_k U_k(\rho_1,\rho_2)}{\partial\rho_1^n\partial\rho_2^m}\bigg|_{\kappa_1,\kappa_2}  \,,
\end{align}
where $\partial_k U_k(\rho_1,\rho_2) $ is given in Eq.~(\ref{eq:uflow}) and initial conditions for Eq.~(\ref{eq:coeflow}) can be transformed from
\begin{align} \label{eq:inifull}
	U_{k=\Lambda}(\rho_1,\rho_2)=a_{10}\rho_1+\frac{1}{2}a_{20}\rho_1^{2}+a_{01}\rho_2 +\Delta U_\Lambda[T,\mu]  \,,
\end{align}
where $\Delta U_\Lambda[T,\mu]$ is the modification of initial conditions. Its form is given in Eq.~(\ref{eq:inimod}), and more detailed descriptions are put in Appendix~\ref{app:inimod}.

With the full effective potential $\Omega_{k}$ as
\begin{align} \label{eq:potenfull}
	\Omega_k(T,\mu,\sigma_L,\sigma_S,L,\bar{L})=\tilde{U}_k(\Sigma,\Sigma^{\dagger})+V_{\rm glue}(L,\bar{L}) \,,
\end{align}
the minimum of the effective potential satisfies
\begin{align}\label{eq:eom}
	\frac{\partial{\Omega_{k=0}}}{\partial \psi} \bigg|_{\psi_p}=0 \,,
\end{align}
where ${\psi}$ represents ($\sigma_L,\sigma_S,L,\bar{L} $) and $\psi_p $ represents the physical solution of $\psi $.

%
\begin{table}[htb]
\caption{Values of parameters for the initial conditions and the Polyakov-loop potential.}
\begin{tabular}{ccccc}
\hline\hline
~~$\Lambda/\rm MeV$~~	&	~~$h $~~	&	~~$a_{10}/\rm {MeV}^2$~~	&	~~$a_{20}$~~	&	~~$a_{01}$~~	\\
\hline
$900 $	&	6.5	&	$830^2 $	&	$10 $	&	$54 $	\\
\hline\hline
$j_l/\rm {MeV}^3$	&	$j_s/\rm {MeV}^3$	&	$c_A/\rm MeV$	&	$\alpha $	&	$b_0 $	\\
\hline
	$120.73^2 $	&	$336.41^3 $	&	$4807.84 $	&	$0.47 $	&	$1.6 $	\\	
\hline\hline		
\end{tabular}
\label{table:parameters}
\end{table}
%

In Table~\ref{table:parameters} we show our used parameters for the initial conditions and the Polyakov-loop potential. The last two are related to the Polyakov-loop potential, see for example Eqs.~(\ref{eq:rescalet}),~(\ref{eq:bmu}), 
while the other parameters are related to the effective action of the matter sector. $\Lambda $ sets our cutoff scale, below which the gluon has developed a mass gap and decoupled from the low energy physics~\cite{Braun:2014ata}. 
The $h,j_{l},j_{s} $ are fixed by the Goldberger-Treiman relation and the PCAC theorem~\cite{Lenaghan:2000ey}. 
The $c_{A} $ represents the strength of the $U_{A}(1) $ anomaly and it depends generally on the external parameters, 
such as the temperature or chemical potential. For the time being we take it as a constant for simplicity. Then one should tune the parameters $a_{10},a_{20},a_{01} $ to produce the vacuum hadronic observables. 
The produced vacuum hadronic observables are listed in Table~\ref{table:masses}, in which $f_{\pi},f_{K} $ can be related to the quark masses via $m_{u,d}=\frac{h}{2}f_{\pi}, m_{s} =\frac{h}{2} (2f_{K} - f_{\pi})$. 
The other important constrains on the parameters come from the pion mass $m_{\pi} =138 \,$MeV, 
the kaon mass $m_{K}=496 \,$MeV, the sum $m_{\eta}^{2} + m_{\eta^\prime}^{2} =1.219\,\rm{GeV}^{2} $, 
and the $\sigma $ meson mass $m_{\sigma} =463\, $MeV. 
In experiments $\sigma $ meson mass shows a large uncertainty, and its value from the latest PDG is $400$--$550 \,$MeV. 
The parameters used here are the same as those used in Refs.~\cite{Fu:2018qsk,Fu:2018swz}.

%
\begin{table}[htb]
\caption{Calculated hadronic observables (in MeV). }
\begin{tabular}{cccccc}
\hline\hline
~~$f_{\pi} $~~ & ~~$f_{K} $~~ & ~~$m_{u,d} $~~ & ~~$m_s $~~ & ~~$m_{a_0} $~~ & ~~$m_{\kappa} $~~ \\
\hline
~~$93 $~~ & $113 $ & $302 $ & $433 $ & $1040 $ & ~~$1139 $~~ \\
\hline\hline
$m_\sigma $ & $m_{f_0} $ & $m_{\pi} $ & $m_{K} $ & $m_{\eta} $ & $m_{\eta^\prime} $ \\
\hline
$463 $ & $1157 $ & $138 $ & $496 $ & $538 $ & $964 $ \\
\hline\hline
\end{tabular}
\label{table:masses}
\end{table}
%

%
\begin{figure*}[htb]
	\includegraphics[width=0.31\textwidth]{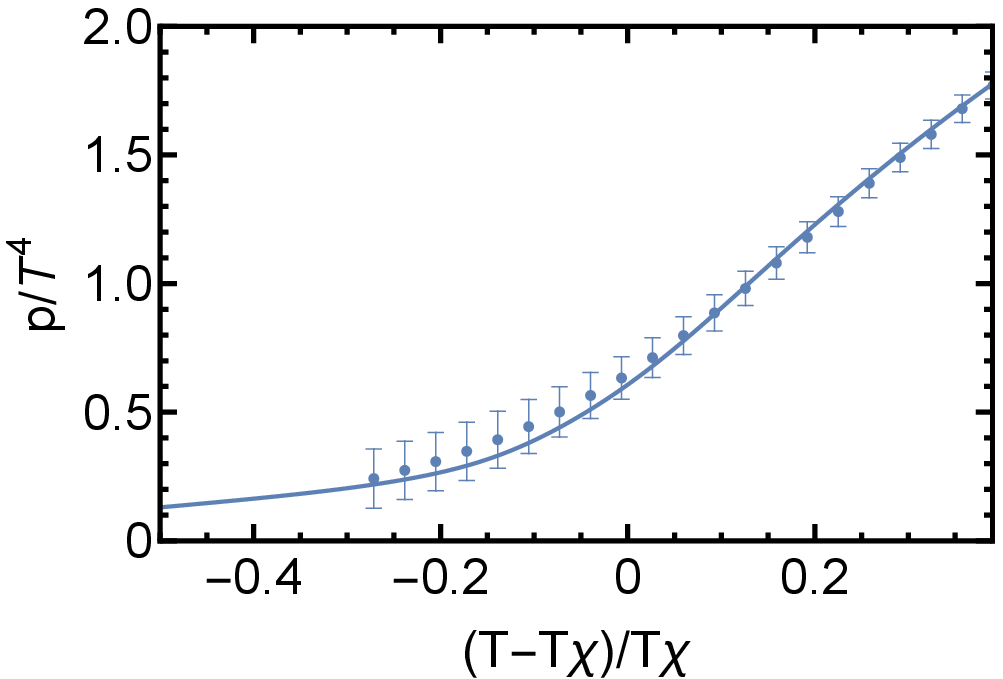} \hspace*{2mm}
	\includegraphics[width=0.295\textwidth]{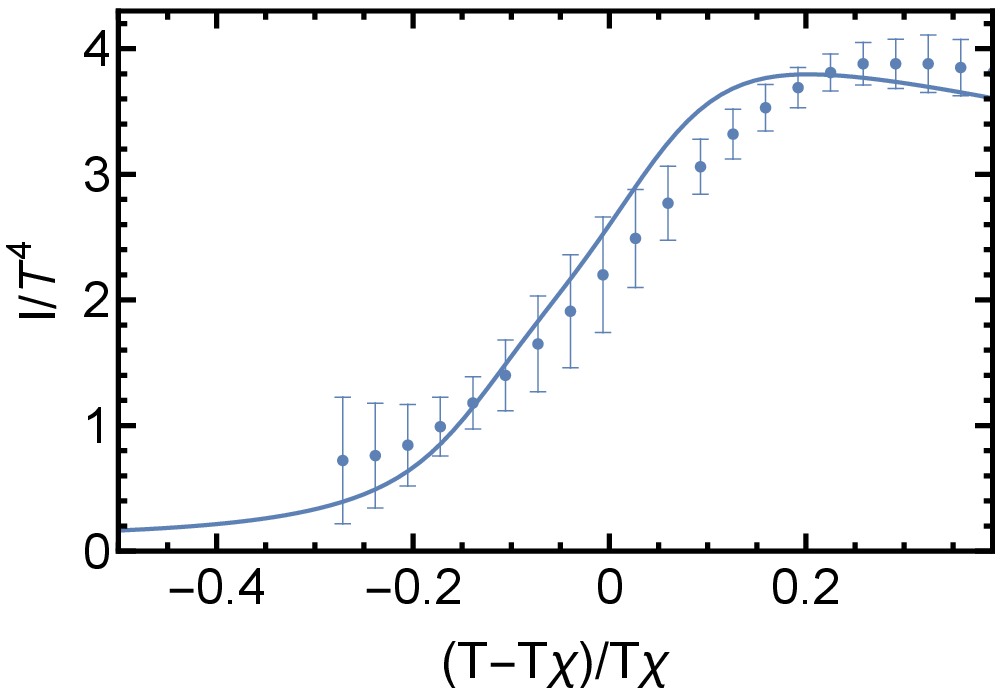} \hspace*{2mm}
	\includegraphics[width=0.31\textwidth]{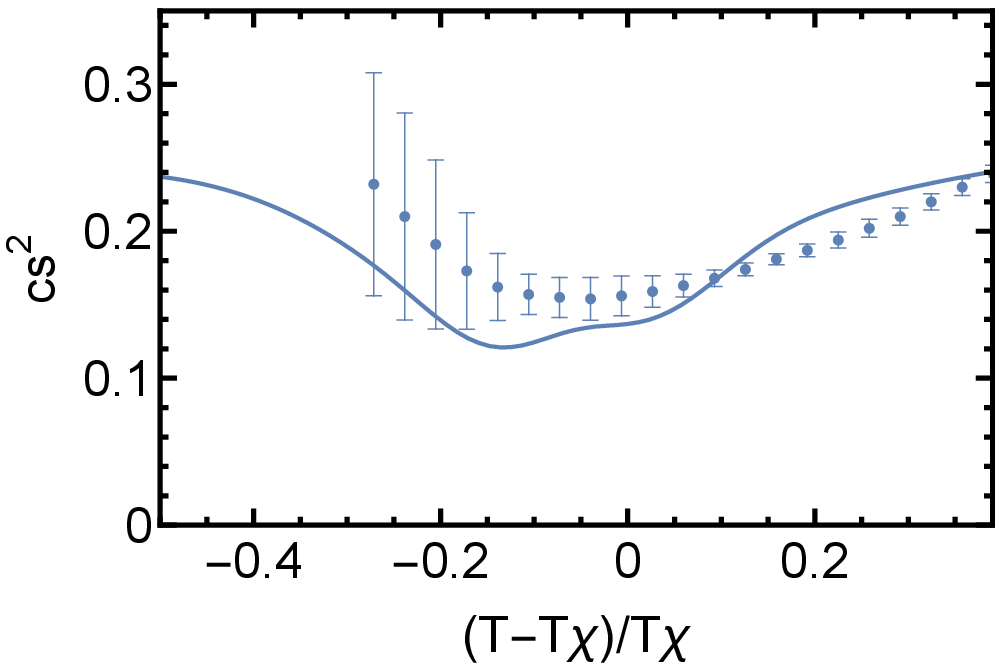}
	\caption{Calculated pressure $p $, trace anomaly $I $ and the speed of sound $c_{s}^{2} $ as functions of the reduced temperature at varnishing density in comparison to the lattice QCD results~\cite{Borsanyi:2013bia}. }
	\label{fig:varnishingmu}
\end{figure*}
%

In Fig.~\ref{fig:varnishingmu}, we show the calculated results of the pressure $p $, the trace anomaly $I\equiv \varepsilon-3p $ and the speed of sound squared $c_{s}^{2}\equiv s/(\partial{\varepsilon}/\partial{T}) $ with $\varepsilon = -p +T\partial{p}/\partial{T} + \mu_{q} n_{q}$ at varnishing density. Since the intrinsic scale of the model is different from that in lattice QCD, $T_{\chi}^{\rm lattice}\sim 155 \,$MeV in lattice QCD simulations~\cite{Borsanyi:2013bia} whereas $T_{\chi}^{\rm model}\sim 176 \,$MeV in this model, 
we compare the results as a function of reduced temperature $t=(T-T_{\chi})/{T_\chi}$. 
One observes a good agreement with the lattice QCD results for the pressure $p $ and the trace anomaly $I $, 
which is not a surprise since we have made use of them to determine the parameters $\alpha,b_{0}$ in the Polyakov-loop potential. 
However it's remarkable that the speed of sound $c_{s}^{2} $ which entails the second order derivatives of the thermodynamical function, also shows a good agreement.

\section{Numerical results}
\label{sec:results}
\subsection{low-order fluctuations and freeze-out parameters}
\label{subsec:low-order}

%
\begin{figure}[htb]
	\includegraphics[width=0.45\textwidth]{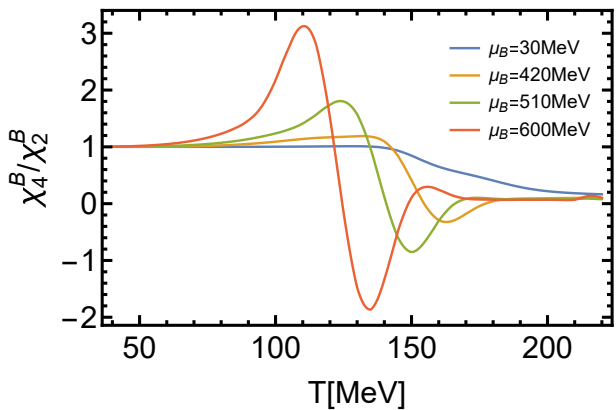}
	\caption{Calculated Ratio of baryon number susceptibilities $\chi_{4}^{B}/\chi_{2}^{B}$ at several values of $\mu_{B}$ in the case of the $\mu$-independent glue potential.}
	\label{fig:R42s}
\end{figure}
%

The baryon number fluctuations are proportional to the powers of the correlation length $\xi$ in the phase transitions, which becomes more prominent when it comes close to the CEP~\cite{Hatta:2003wn,Stephanov:2011pb}. We can see in Fig.~\ref{fig:R42s} that fluctuations becomes more and more significant with the increase of $\mu_{B}$. Thus one can investigate the phase transitions by studying the baryon number susceptibilities. The theoretical susceptibility ratios are related to moments of the multiplicity distributions of the conserved charges in experiments by
\begin{align}
	\frac{\chi_{2}^{B}}{\chi_{1}^{B}} = \sigma^{2}/M \,, \quad \frac{ \chi_{3}^{B}}{\chi_{2}^{B}} = S\sigma \,, \quad \frac{\chi_{4}^{B}}{\chi_{2}^{B}} = \kappa \sigma^{2} \,,
\end{align}
where $\sigma^{2},M,S $ and $\kappa $ are the invariance, mean,  skewness and kurtosis of the multiplicity distributions, respectively.
The first susceptibility $\chi_1^B$ is related to the baryon number density by $\chi_1^B=n_B/T^3$. The particle number density can be obtained via
\begin{align}
\frac{d\Omega_{k=0}}{d\mu}=\frac{\partial\Omega_{k=0}}{\partial\mu}+\frac{\partial\Omega_{k=0}}{\partial\psi}\frac{\partial\psi}{\partial\mu}=\frac{\partial\Omega_{k=0}}{\partial\mu}\,,
\end{align}
where the equations of motion, Eq.~(\ref{eq:eom}), are used in the last equality. Therefore, we partially differentiate Eqs.~(\ref{eq:uexpand}),~(\ref{eq:coeflow}) to compute $\chi_{1}^{B}$.
Subsequently, the higher order susceptibilities can be obtained by numerical differentiation of $\chi_{1}^{B}$,
\begin{align}
	\chi_{i+1}^{B} &= \frac{1}{\beta^{i}} \frac{\partial^{i}{\chi_1^B}}{\partial{\mu_B^i}}\,,
\end{align}
where $\beta=1/T,i=1,2,3\cdots$.

%
\begin{figure*}[t!]
	\includegraphics[width=0.44\textwidth]{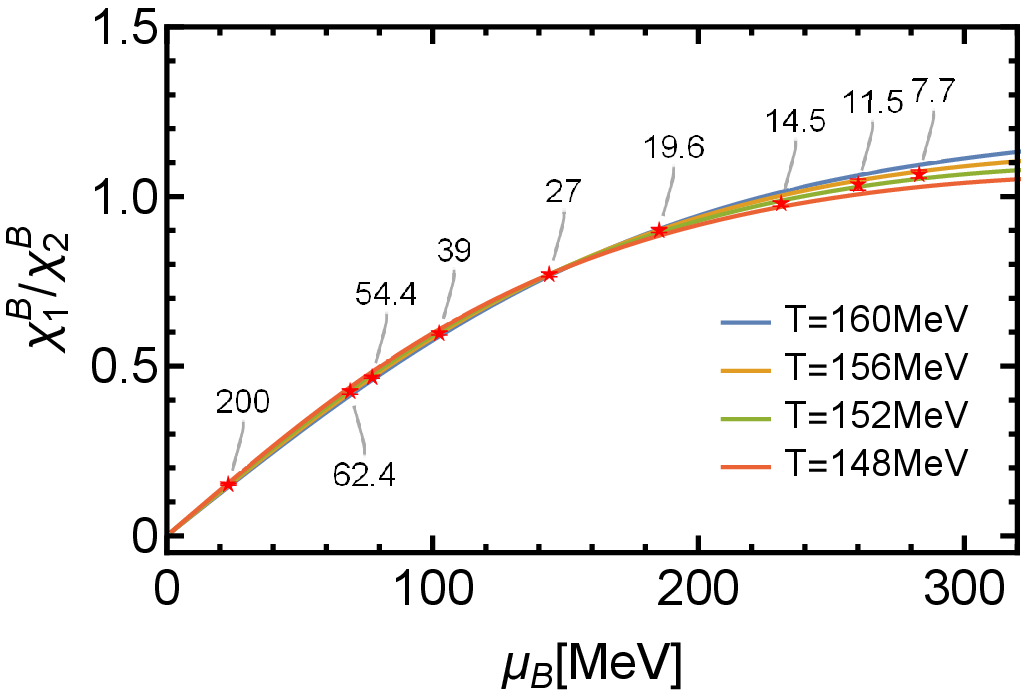} \hspace*{5mm}
	\includegraphics[width=0.44\textwidth]{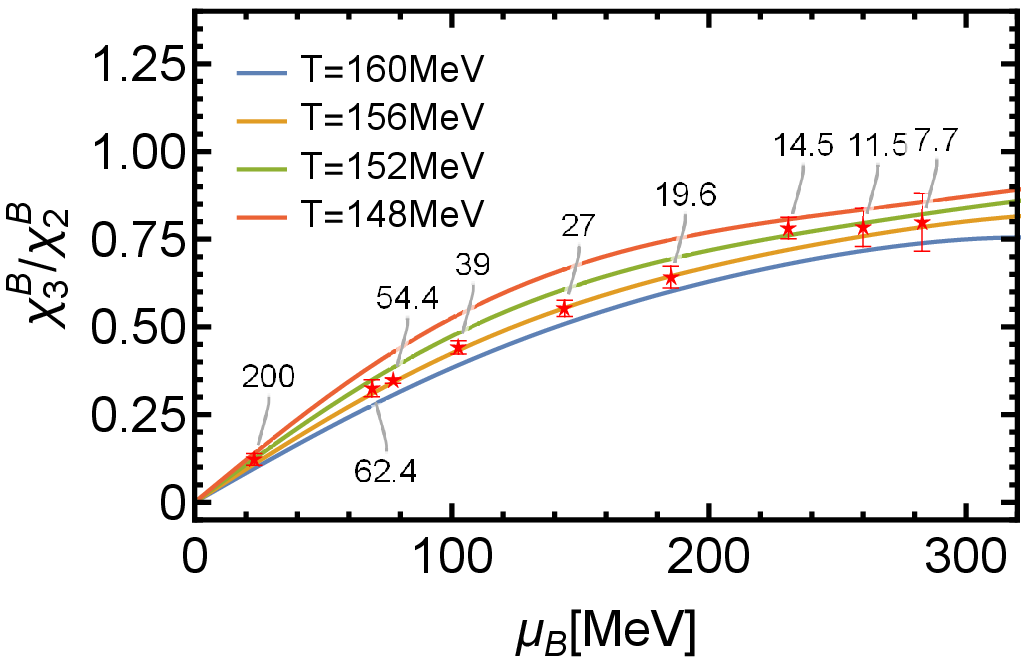}
	\caption{
		Calculated baryon chemical potential dependence of the susceptibility ratios $\chi_{1}^{B}/\chi_{2}^{B}$ (left panel) and $\chi_{3}^{B}/\chi_{2}^{B}$ (right panel) at different temperatures in the case of $\mu$-dependent glue potential. The red stars stand for our assigned freeze-out points for the center values of the experimental data~\cite{STAR:2020tga} at collision energies $\sqrt{S_{NN}}=200, 62.4, 54.4, 39, 27, 19.6, 14.5, 11.5, 7.7\,$GeV. Error bars in the red stars represent the experimental errors of $\chi_{1}^{B}/\chi_{2}^{B}$ and $\chi_{3}^{B}/\chi_{2}^{B}$ at different collision energies. }
	\label{fig:R12&R32mub}
\end{figure*}
%

%
\begin{table*}[htb]
	\renewcommand{\arraystretch}{1.3}
	\caption{Obtained freeze-out points $(\mu_B^f, T^f)$ for $\mu$-dependent and $\mu$-independent glue potentials ($(\mu_B^f, T^f)$ are in unit MeV and $\sqrt{S_{NN}}$ in GeV). }
	\label{table:freeze-out points}
	\begin{tabular}{|cc|c|c|c|c|c|c|c|c|c|}
		\hline
		\multicolumn{2}{|c|}{$\sqrt{S_{NN}}$}                                & $200$ & $62.4$ & $54.4$ & $39$ & $27$ & $19.6$ & $14.5$ & $11.5$ & $7.7$ \\ \hline
		\multicolumn{1}{|c|}{\multirow{2}{*}{$\mu$-dependent}}   & $\mu_B^f$ 	& $23.2_{-0.6}^{+0.6}$ & $68.9_{-0.8}^{+0.8}$ & $77.1_{-0.2}^{+0.2}$ & $102.6_{-0.4}^{+0.4}$ & $143.9_{-0.2}^{+0.3}$ & $185.2_{-0.7}^{+1.1}$ & $231._{-5.3}^{+6.1}$ & $260.0_{-9.9}^{+14.2}$ & $282.9_{-19.6}^{+37.5}$      \\ \cline{2-11}
		\multicolumn{1}{|c|}{}                                   & $T^f$    & $152.5_{-4.2}^{+4.2}$ & $154.3_{-2.2}^{+2.3}$ & $155.5_{-0.5}^{+0.5}$ & $155.3_{-1.0}^{+1.1}$ & $156.1_{-1.4}^{+1.6}$ & $156.2_{-2.0}^{+2.2}$ & $150.1_{-1.9}^{+2.0}$ & $153.3_{-2.9}^{+3.1}$ & $154.6_{-4.3}^{+5.2}$      \\ \hline
		\multicolumn{1}{|c|}{\multirow{2}{*}{$\mu$-independent}} & $\mu_B^f$	& $23.1_{-0.6}^{+0.6}$ & $68.6_{-0.7}^{+0.8}$ & $76.7_{-0.2}^{+0.2}$ & $102.0_{-0.3}^{+0.3}$ & $142.8_{-0.0}^{+0.1}$ & $183.8_{-1.1}^{+1.4}$ & $231.2_{-5.4}^{+6.4}$ & $262.9_{-12.4}^{+16.9}$ & $288.4_{-26.5}^{+50.2}$        \\ \cline{2-11}
		\multicolumn{1}{|c|}{}                                   & $T^f$	& $151.9_{-4.2}^{+4.0}$,& $153.6_{-2.1}^{+2.2}$,& $154.6_{-0.5}^{+0.5}$,& $154.3_{-1.0}^{+1.0}$,& $154.6_{-1.3}^{+1.4}$,& $154.1_{-1.7}^{+1.9}$,& $147.9_{-1.8}^{+1.8}$,& $149.8_{-2.8}^{+3.0}$,& $150.3_{-4.7}^{+5.3}$       \\ \hline
	\end{tabular}
	\renewcommand{\arraystretch}{1.0}
\end{table*}
%

%
\begin{table}[htb]
	\renewcommand{\arraystretch}{1.1}
	\caption{Parameters $T_0, a, b$ in Eq.~(\ref{eq:fitmodel-tmu}) and $c, d$ in Eq.~(\ref{eq:fitmodel-mue}) for both $\mu$- dependent and independent $T_0$ cases ($c, T_0^f$ are in unit MeV and $d$ in $\mathrm{GeV}^{-1}$). }
	\label{table:freeze-out pars}
	\begin{tabular}{cccccc}
		\hline\hline
		~~$T_0$~~	&	~~$c$~~	&	~~$d$~~	&	~~$T_{0}^{f}$~~	&	~~$a$~~		& $b$	\\
		\hline
		$\mu$-dependent	&	$708.96$	&	$0.147$	&	$154.8$	&	$4.05\times10^{-7}$	& $0.0022$\\
		\hline
		$\mu$-independent	& 	$751.31$	&	$0.159$	&	$153.9$	&	$7.35\times10^{-7}$	& $0.0040$	\\
		\hline\hline		
	\end{tabular}
	\renewcommand{\arraystretch}{1.0}
\end{table}
%

%
\begin{figure*}[htb]
	\includegraphics[width=0.44\textwidth]{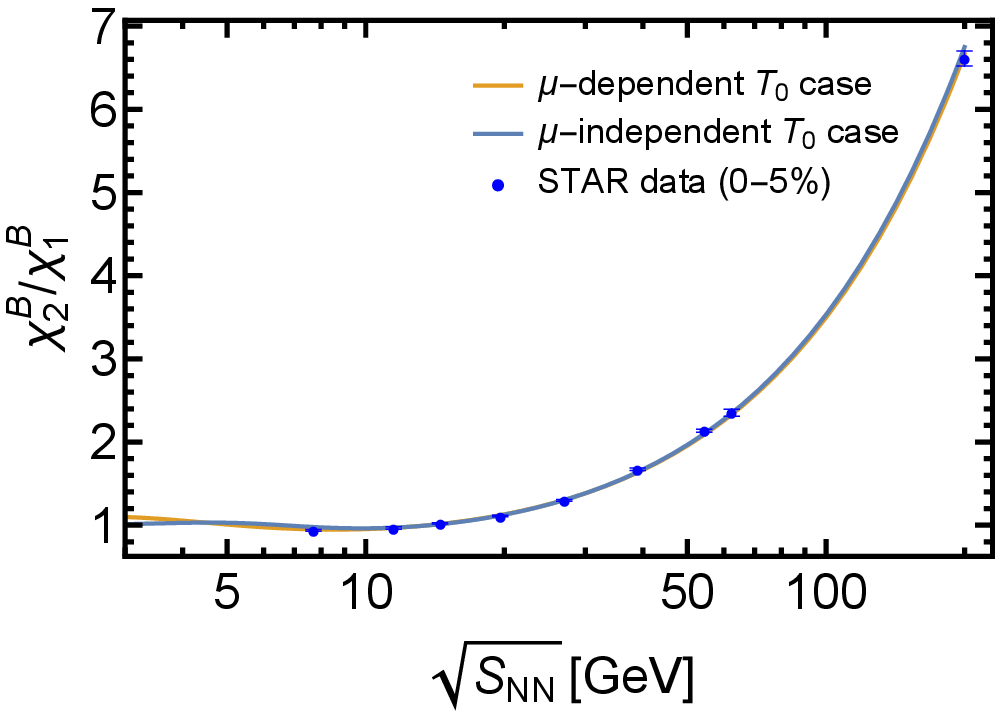} \hspace*{5mm}
	\includegraphics[width=0.45\textwidth]{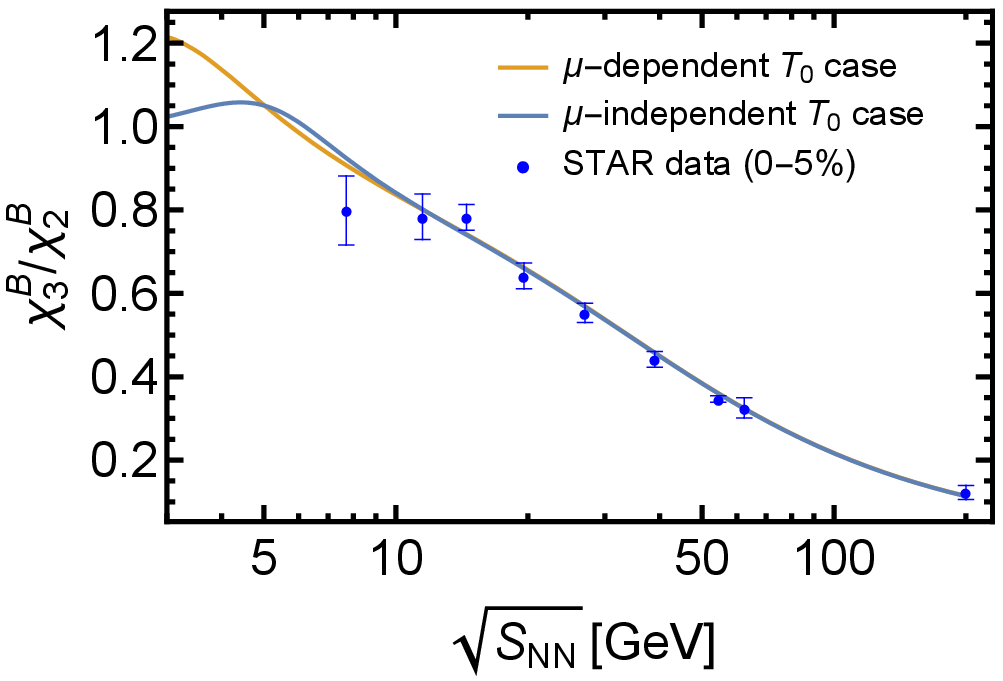}
	\caption{Comparison of the calculated $\chi_{2}^{B}/\chi_{1}^{B}, \chi_{3}^{B}/\chi_{2}^{B}$ for both $T_{0}$ cases along the respective freeze-out line with the experimental data~\cite{STAR:2020tga}.}
	\label{fig:compareR21&R32}
\end{figure*}
%

Then by comparing the theoretically calculated ratios $\chi_{2}^{B}/\chi_{1}^{B}$ and $\chi_{3}^{B}/\chi_{2}^{B}$ with the experimental data we can determine the freeze-out points $(\mu_{B}^{f}, T^{f})$~\cite{Alba:2014eba,Bazavov:2012vg,Borsanyi:2014ewa,Karsch:2012wm}. 
In Fig.~\ref{fig:R12&R32mub}, we illustrate the obtained $\mu_{B}$ dependence of $\chi_{1}^{B}/\chi_{2}^{B}$ and $\chi_{3}^{B}/\chi_{2}^{B}$ at different temperatures in the case of $\mu$-dependent glue potential, and show our assigned freeze-out points for center values of the experimental data~\cite{STAR:2020tga} at different collision energies. Note that the error bars in the figure represent the experimental error of $\chi_{1}^{B}/\chi_{2}^{B}$ and $\chi_{3}^{B}/\chi_{2}^{B}$ at different collision energies. 
As the experimental data contains systematic and statistical errors, we use the statistical error to change the experimental data and then estimate the influence on the change of the freeze-out points. 
In this way we obtain the freeze-out points $(\mu_{B}^{f}, T^{f})$ with errors stemming from the the experimental errors at different collision energies. The obtained results are listed in Table~\ref{table:freeze-out points}.

With the obtained freeze-out points, the freeze-out line is fitted as
\begin{align}
	T^{f} &  =  T_{0}^{f} \Big[ 1 - a \Big( \frac{\mu_{B}^{f}}{T_{0}^{f}} \Big)^2
	- b \Big( \frac{\mu_{B}^{f}}{T_{0}^{f}} \Big)^4 \Big] \, \label{eq:fitmodel-tmu} \\
	\mu_{B}^{f} & =  \frac{c}{1+d \sqrt{S_{NN}^{}} } \, . \label{eq:fitmodel-mue}
\end{align}
Since the last data corresponding to $\sqrt{S_{NN}}=7.7\,$GeV in Table~\ref{table:freeze-out points} shows large uncertainty, we do not use it for fitting the parameters. The best-fitted parameters are listed in Table~\ref{table:freeze-out pars}. With the obtained parameters we can use Eqs.~(\ref{eq:fitmodel-tmu}),~(\ref{eq:fitmodel-mue}) to predict the freeze-out points $(\mu_{B}^{f},T^f)$ of the system at any collision energy.

We illustrate the calculated $\chi_{2}^{B}/\chi_{1}^{B}, \chi_{3}^{B}/\chi_{2}^{B}$ along the respective freeze-out lines as a function of collision energy $\sqrt{S_{NN}}$ for $\mu$-dependent and independent cases, and the comparison with the experimental data in Fig.~\ref{fig:compareR21&R32}. 
One can see a good agreement with the experimental data except for the collision energy $\sqrt{S_{NN}}\sim 7.7\,$GeV. 
On one hand we do not include the last experimental data corresponding to $\sqrt{S_{NN}}=7.7\,$GeV in the fitting process due to its large uncertainty,
on the other hand the plateau of $\chi_{3}^{B}/\chi_{2}^{B}$ below $\sqrt{S_{NN}}\sim14.5\,$GeV may indicate the unaccounted effects in our work, 
for example the identification of baryon number fluctuations with the proton number fluctuations, volume fluctuations, global baryon number conservation at low collision energy~\cite{Braun-Munzinger:2020jbk}, and so forth. 
It should also be noted that, since we take the $\chi_{2}^{B}/\chi_{1}^{B}$ and $\chi_{3}^{B}/\chi_{2}^{B}$ to determine our freeze-out points listed in Table~\ref{table:freeze-out points}, and then these points are used to produce the freeze-out line by fitting Eqs.~(\ref{eq:fitmodel-tmu}),~(\ref{eq:fitmodel-mue}), the obtained parameters are listed in Table~\ref{table:freeze-out pars}. 
The degree of the agreement between theoretical calculation results and experimental data shown in Fig.~\ref{fig:compareR21&R32} can also be viewed as a test of our fitting process. 
Moreover, we can see that the lines for both $\mu$-dependent and independent $T_{0}$ cases coincide almost with each other for $\sqrt{S_{NN}}>7.7\,$GeV. 
This is due to the slight difference in low order susceptibilities between two $T_{0}$ cases in the region of $\mu_{B}$ being not high. 
However, because of the direct $\mu$ dependence in the glue potential, the difference in the higher order fluctuations between the two cases will be more apparent, see for example Fig.~\ref{fig:compareR62}.

The parameter $a$ describes the curvature of the freeze-out line at large collision energy, and its small value in our fitting process is closely related to the slightly increasing trend of the temperature of the freeze-out points with $\mu_{B}$ at large $\sqrt{S_{NN}}$, see the freeze-out points plotted in Fig.~\ref{fig:phasediagramC&F} at small $\mu_B$. The trend also appears in Ref.~\cite{Alba:2014eba}, and small values of the parameter $a$ are consistent with the conclusions in Ref.~\cite{Bazavov:2015zja}.

\subsection{Freeze-out lines and high-order fluctuations}
\label{subsec:high-order}
%
\begin{figure}[htb]
	\includegraphics[width=0.47\textwidth]{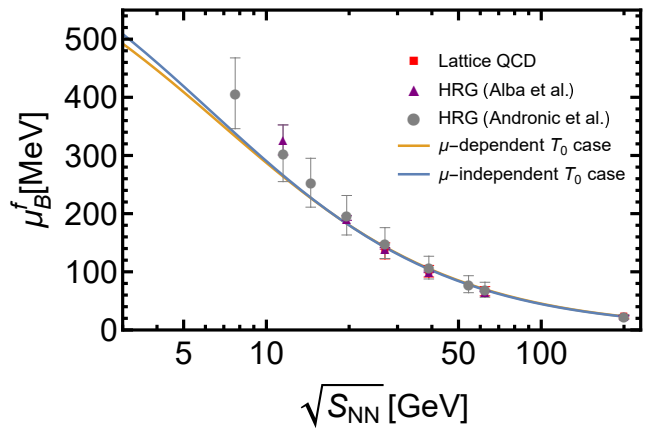}
	\caption{Comparison of the collision energy dependence of the freeze-out chemical potential $\mu_{B}^{f}(\sqrt{S_{NN}})$ with those given in the lattice QCD simulations~\cite{Borsanyi:2014ewa}, HRG model (Alba et al.)~\cite{Alba:2014eba} and HRG model (Andronic et al.)~\cite{Andronic:2005yp,Andronic:2017pug}. }
	\label{fig:comparemue}
\end{figure}
%

%
\begin{figure*}[htb]
	\includegraphics[width=0.44\textwidth]{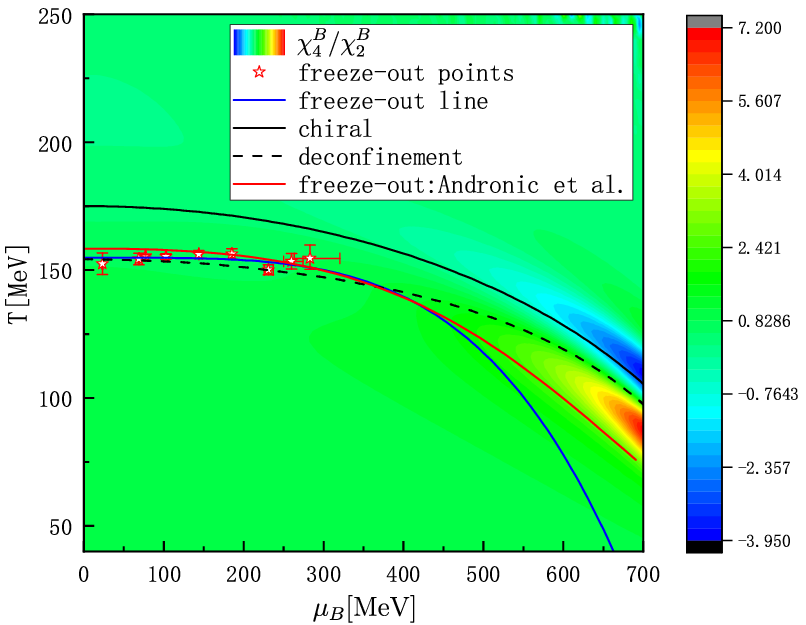} \hspace*{2mm}
	\includegraphics[width=0.44\textwidth]{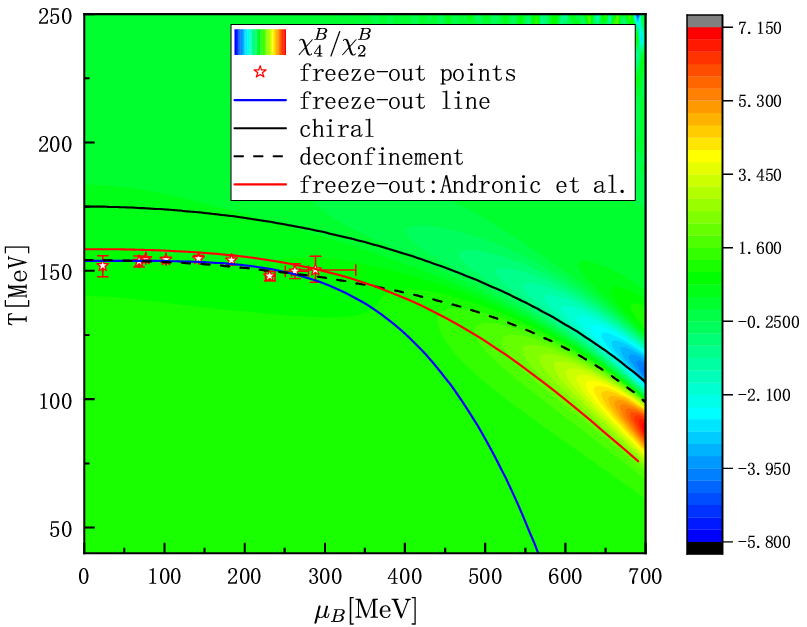}
	\caption{Calculated phase diagrams for $\mu$-dependent and $\mu$-independent $T_0$ cases. The left panel is for the  $\mu$-dependent case, 
while the right panel is for the $\mu$-independent case. The background is the density plot of the susceptibilities ratio $R^{B}_{42}\equiv\chi_{4}^{B}/\chi_{2}^{B}$. Red stars are the obtained freeze-out points in the respective cases, see Table~\ref{table:freeze-out points} for their specific values. Solid blue, solid black and dashed black curves represent the freeze-out, the chiral phase transition and the deconfinement phase transition lines, respectively. Solid red curve is the freeze-out line from Andronic {\it et al.}~\cite{Andronic:2017pug}.     }
	\label{fig:phasediagramC&F}
\end{figure*}
%

%
\begin{figure*}[htb]
	\includegraphics[width=0.42\textwidth]{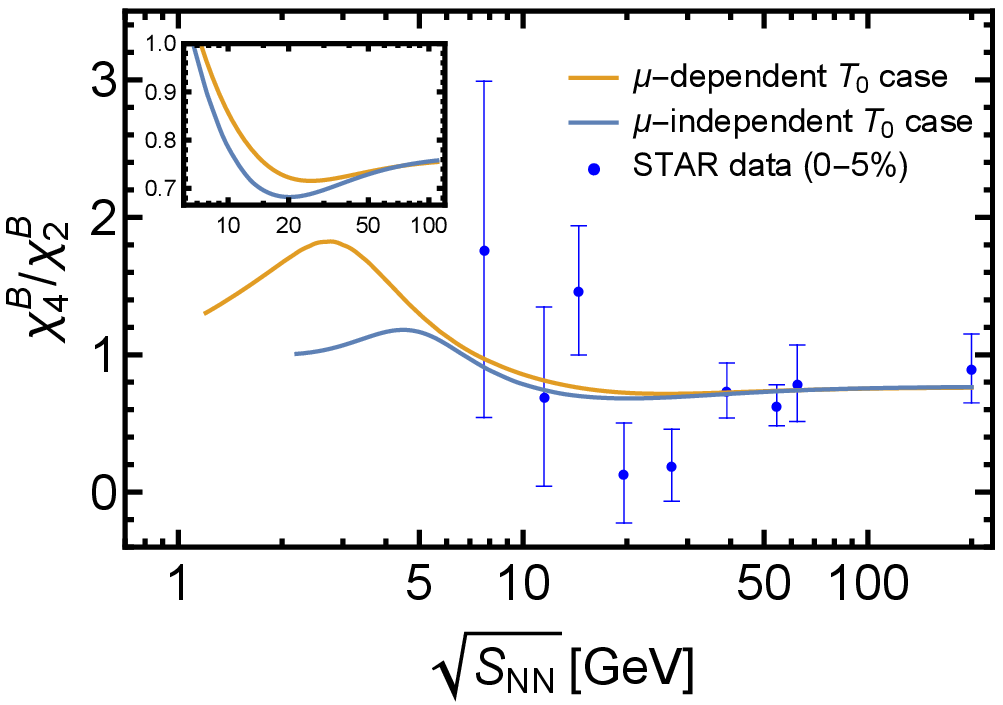} \hspace*{5mm}
	\includegraphics[width=0.46\textwidth]{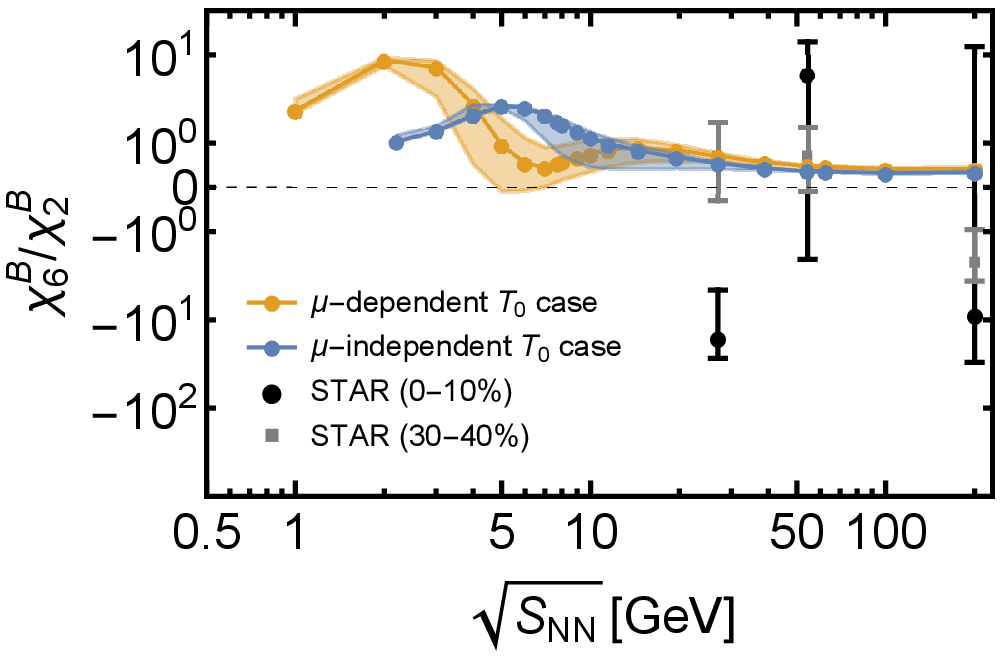}
	\caption{Left panel: Calculated collision energy dependence of  $\kappa\sigma^{2}=\chi_{4}^{B}/\chi_{2}^{B}$ along the freeze-out line. 
The blue circles are the experimental values~\cite{STAR:2020tga} for $0\%-5\%$ centrality. The inlay zooms in at $\sqrt{S_{NN}}\sim 20\,$GeV. \\
	Right panel: Calculated collision energy dependence of $\chi_{6}^{B}/\chi_{2}^{B}$ (in logarithmic scale) along the freeze-out line. The shadowed regions represent our numerical uncertainties. The black circles and gray squares are the experimental values~\cite{STAR:2021rls} for $0\%-10\%$ and $30\%-40\%$ centralities, respectively.   }
	\label{fig:compareR42}
	\label{fig:compareR62}
\end{figure*}
%

With the obtained parameters in Eqs.~(\ref{eq:fitmodel-tmu}) and (\ref{eq:fitmodel-mue}), one can predict the higher order fluctuations along the freeze-out line as a function of collision energy $\sqrt{S_{NN}}$. To this end, we first illustrate the calculated dependence of the freeze-out baryon chemical potential $\mu_{B}^{f}$ on the collision energy $\sqrt{S_{NN}}$, and the comparison with those given in the lattice QCD simulations~\cite{Borsanyi:2014ewa}, and other model calculations~\cite{Alba:2014eba,Andronic:2005yp,Andronic:2017pug} in Fig.~\ref{fig:comparemue}. 
We see that the freeze-out chemical potential matches the lattice QCD and other model results well for $\mu_{B} \lesssim 300\,$MeV, while for $\mu_{B} \gtrsim 300\,$MeV it begins to deviate from the HRG results.

We also depict the freeze-out lines and that from Andronic {\it et al.}~\cite{Andronic:2017pug} in Fig.~\ref{fig:phasediagramC&F}, 
as well as the phase transition lines and the obtained freeze-out points in the two $T_0$ cases. 
The background in Fig.~\ref{fig:phasediagramC&F} is the density plot of the susceptibility ratio $R^{B}_{42}\equiv\chi_{4}^{B}/\chi_{2}^{B}$. 
The chiral phase transition point is defined as the inflection point of the subtracted chiral condensate $\Delta_{LS}$,
\begin{align}\label{eq:subcond}
	\Delta_{LS} = \frac{\big(\sigma_L - \frac{j_l}{j_s} \sigma_S\big)\big|_{T}}{\big(\sigma_L - \frac{j_l}{j_s} \sigma_S\big)\big|_{T=0}} \,,
\end{align}
while the deconfinement phase transition point is defined as the inflection point of the Polyakov-loop $L$. 
Their respective values at vanishing density are $T^{d}\sim 155\,$MeV and $T^{\chi} \sim 176\,$MeV. 
The position of CEP is not resolved in the model since our expansion is not fully convergent for $\mu_{B}>700\,$MeV. 
One can see that the freeze-out line in the $\mu$-dependent case tends to larger $\mu_{B}$ for the low temperature, 
and is overall closer to that from Andronic {\it et al.}~\cite{Andronic:2017pug} when compared to the $\mu$-independent case.

Then we calculate the $\chi_4^B/\chi_2^B$ along the freeze-out line as a function of the collision energy $\sqrt{S_{NN}}$ for both $T_{0}$ cases. 
They are depicted in the left panel of Fig.~\ref{fig:compareR42}. 
The inlay zooms in at $\sqrt{S_{NN}}\sim 20\,$GeV and indeed shows a nonmonotonicity of $\chi_{4}^{B}/\chi_{2}^{B}$, 
which is somehow rather weak compared to the STAR data ($0-5\%$). 
When $\sqrt{S_{NN}}$ comes to lower, $\chi_{4}^{B}/\chi_{2}^{B}$ in Fig.~\ref{fig:compareR42} shows a peak and then approaches to one in both $T_{0}$ cases. The peak positions and their heights are different for the two cases.

We further calculate the $\sqrt{S_{NN}}$ dependence of $\chi_{6}^{B}/\chi_{2}^{B}$ along the obtained freeze-out lines. 
The obtained results are depicted in the right panel of Fig.~\ref{fig:compareR62}. 
We take bands to represent our numerical errors for  $\chi_{6}^{B}/\chi_{2}^{B}$. 
For the $\mu$-dependent case, with the decreasing collision energy, $\chi_{6}^{B}/\chi_{2}^{B}$ begins to have a minimum around $\sqrt{S_{NN}}\sim 5-10\,$GeV, and then have a maximum around $\sqrt{S_{NN}}\sim 2\,$GeV, and approaches to one for the lower collision energy. 
The $\mu$-independent case shows a similar behavior, except that the minimum becomes very flat.


It should also be mentioned that there are lots of things to be included for an improvement on the current calculations and results. The LPA approximation is employed in the present calculation, while it is shown that the baryon number fluctuations calculated beyond the LPA approximation could be more prominent~\cite{Fu:2015naa}. Constrains from charge conservation such as the baryon number conservation, the strangeness neutrality and the electric charge conservation should be considered for a more realistic setting of the experimental situation, which may play a important role especially at large density. Also the finite volume effects can make a sizable difference on the baryon number susceptibilities~\cite{Almasi:2016zqf,Skokov:2012ds,Lu:2021ium}. Last but not least, by the inclusion of the gluon and ghost fields via the dynamical hadronization one can advance the low energy effective model towards full QCD further, and the progress has been made in Ref.~\cite{Fu:2019hdw}. With all these improvements one may obtain a better freeze-out line in Fig.~\ref{fig:phasediagramC&F}, and the agreement of collision energy dependence of the freeze-out chemical potential with other models in Fig.~\ref{fig:comparemue} would be improved especially at low collision energy. The weak nonmonotonicity shown in Fig.~\ref{fig:compareR42} may also be overcomed. The relevant improvements will be done in the future.

\section{Summary}
\label{sec:summary}

In summary, we have calculated the baryon number susceptibilities with the FRG approach which incorporates the non-perturbative quantum, thermal and density fluctuations. For a better treatment of finite density situations we use the hard thermal or density loop improved $\mu$-dependent Polyakov-loop potential, and compare the results with those obtained from $\mu$-independent Polyakov-loop potential in the main text. We calculate the ratios $\chi_{2}^{B}/\chi_{1}^{B}$ and $\chi_{3}^{B}/\chi_{2}^{B}$ at different baryon chemical potentials and temperatures, 
and by comparing them with the experimental data we determine the freeze-out points $(\mu_{B}^{f},T^{f})$ at different collision energies. 
The obtained collision energy dependence of $\chi_{2}^{B}/\chi_{1}^{B}$ and $\chi_{3}^{B}/\chi_{2}^{B}$ are in good agreement with the experiment, 
see for example Fig.~\ref{fig:compareR21&R32}. With the obtained freeze-out points we fit the freeze-out line via fitting model Eqs.~(\ref{eq:fitmodel-tmu}) and (\ref{eq:fitmodel-mue}), and the collision energy dependence of the freeze-out chemical potentials agrees well with the lattice QCD simulations and other model calculations for $\sqrt{S_{NN}}>7.7\,$GeV. 
The freeze-out line obtained with the hard thermal or density loop improved $\mu$-dependent Polyakov-loop potential is closer to that from Andronic {\it et al.}~\cite{Andronic:2017pug} when compared to that with $\mu$-independent Polyakov-loop potential. 
Then we calculate the $\chi_{4}^{B}/\chi_{2}^{B}$ and $\chi_{6}^{B}/\chi_{2}^{B}$ along the respective freeze-out lines. 
The obtained $\chi_{4}^{B}/\chi_{2}^{B}$ shows a weak nonmonotonicity around $\sqrt{S_{NN}}\sim 20\,$GeV in both $\mu$-dependent and independent $T_{0}$ cases, which then develops a maximum and approaches to one with the decreasing collision energy. 
The obtained $\chi_{6}^{B}/\chi_{2}^{B}$ shows a similar complicated behavior, which develops a minimum and then a maximum with the decreasing collision energy, and approaches to one for the lower collision energy.


\begin{acknowledgments}
The work was supported by the National Natural Science Foundation of China under Grant Nos. 12175007, 12175030, and 12247107. 
%
\end{acknowledgments}

\appendix

\section{The Polyakov-loop potential and threshold functions}\label{app:eqs}

The Polyakov-loop potential $V_{\rm glue}$ in Eq.~(\ref{eq:Vglue}) is
\begin{align}
	\frac{V_{\rm glue}(L,\bar L)}{T^4} &= -\frac{a(T)}{2} \bar L L +  b(T)\ln M_H(L,\bar{L}) \nonumber\\
	&\quad + \frac{c(T)}{2} (L^3+\bar L^3) + d(T) (\bar{L} L)^2 \,,
\end{align}
with the Haar measure $M_H(L,\bar{L})$
\begin{align} \label{eq:haar}
	M_H (L, \bar{L})&= 1 -6 \bar{L}L + 4 (L^3+\bar{L}^3) - 3 (\bar{L}L)^2 \,.
\end{align}
The coefficients in Eq.~(\ref{eq:Vglue}) is of the form
\begin{align}\label{eq:xT}
	x(T) = \frac{x_1 + x_2/(t+1) + x_3/(t+1)^2}{1 + x_4/(t+1) + x_5/(t+1)^2}
\end{align}
for $x\in \{a,c,d\} $, and
\begin{align}\label{eq:bT}
	b(T) =b_1 (t+1)^{- b_4}\left (1 -e^{b_2/(t+1)^{b_3}} \right) \,,
\end{align}
with
\begin{align} \label{eq:tYM}
	t=(T-T_{\rm YM})/T_{\rm YM}\,,
\end{align}
where $T_{\rm YM}$ is the critical temperature for the deconfinement phase transition in the pure gauge theory. Values of parameters in Eqs.~(\ref{eq:xT}) and (\ref{eq:bT}) are collected in Table~\ref{table:coeffs}.
\begin{table}[htb!]
	\caption{Values of parameters in Eqs.~(\ref{eq:xT}) and (\ref{eq:bT}).}
	\label{table:coeffs}
	\centering
	\begin{tabular}{cccccc}
		\hline\hline
		& 1 & 2 & 3 & 4 & 5 \\
		\hline
		$a_i$ &-44.14& 151.4 & -90.0677 &2.77173 &3.56403 \\
		$b_i$ &-0.32665 &-82.9823 &3.0 &5.85559  &\\
		$c_i$ &-50.7961 &114.038 &-89.4596 &3.08718 &6.72812\\
		$d_i$ &27.0885 &-56.0859 &71.2225 &2.9715 &6.61433\\
		\hline\hline
	\end{tabular}
\end{table}

The threshold functions $l_{0}^{(B/F)} $ in Eq.~(\ref{eq:uflow}) read
\begin{align}
	l_{0}^{(B)}(\bar{m}^2,T,\mu)&=\frac{1}{3\sqrt{1+\bar{m}^{2}}}\Big(1+n_{B}(\bar{m}^{2},T,\mu)\nonumber\\
	&\quad +n_{B}(\bar{m}^{2},T,-\mu)\Big) \,, \nonumber\\
	l_{0}^{(F)}(\bar{m}^2,T,\mu)&=\frac{1}{3\sqrt{1+\bar{m}^{2}}}\Big(1-n_{F}(\bar{m}^{2},T,\mu,L,\bar L)\nonumber\\
	&\quad -n_{F}(\bar{m}^{2},T,-\mu,\bar L,L)\Big) \,,
\end{align}
with
\begin{align}
	n_{B}(\bar{m}^{2};T,\mu)=\frac{1}{e^{(E-\mu)/T}-1} \,,
\end{align}
\begin{align} \label{eq:nF}
	&n_{F}(\bar{m}^{2};T,\mu,L,\bar L) \nonumber\\
	&=\frac{1+2\bar L\,e^{(E-\mu)/T}+L\,e^{2(E-\mu)/T}}{1+3\bar L\,e^{(E-\mu)/T}+3L\, e^{2(E-\mu)/T}+e^{3(E-\mu)/T}} \,,
\end{align}
where $E=k\sqrt{1+\bar{m}^2} $ is the particle energy. Eq.~(\ref{eq:nF}) is the Polyakov-loop modified fermion distribution function, which has an intuitive interpretation. With $L,\bar{L}=0 $, one has $n_{F}\sim 1/(1+e^{3(E-\mu)/T}) $, which is distribution function for a qqq-state. At low temperature, the modified fermion distribution function can be related to the correct counting of baryonic degrees of freedom in a subtle manner, see Ref.~\cite{Fu:2015naa} for a detailed discussion. At high temperature $L,\bar{L}\sim 1 $, one recovers the ordinary fermion distribution function $n_{F}\sim 1/(1+e^{(E-\mu)/T}) $.

\section{Modification of initial conditions}\label{app:inimod}

When the external parameters of the temperature and chemical potential, such as $2\pi T $ and  $2\pi\mu $, that represent additional energy scales, are comparable to the cutoff scale $\Lambda $ of the model, the initial conditions $\Gamma_{k=\Lambda} $ would develop a dependence on these parameters. We follow the procedure proposed in Refs.~\cite{Fu:2018qsk,Wen:2018nkn}, and for more detailed discussions see also Refs.~\cite{Braun:2003ii,Braun:2018svj}. The basic idea is to start the flow from a high enough scale in which the dependence of the external parameters can be neglected, and flow down to the cutoff scale, and then the modification of initial conditions is
\begin{align} \label{eq:inimodg}																										
	\Delta\Gamma_{\Lambda}[s]\ &=\Gamma_{\Lambda}(s) - \Gamma_{\Lambda}(0)  \nonumber\\
	&=[\Gamma_{\Lambda}(s) - \Gamma_{\infty}(s)] - [\Gamma_{\Lambda}(0) - \Gamma_{\infty} (0)] \nonumber\\
	&=\int^{\Lambda}_{\infty} dk\ [\partial_{k} \Gamma_{k}(s) - \partial_{k}\Gamma_{k}(0)] \,,
\end{align}	
with $s $ denotes all the external parameters ($T $, $\mu $ in the present case). Since quark fluctuations dominate over meson fluctuations at high energy scale, we could approximate the flow in Eq.~(\ref{eq:inimodg}) by fermionic parts, to wit,
\begin{align}\label{eq:inimod}
	&\Delta U_\Lambda[T,\mu]= \int_\infty^{\Lambda} \!dk\frac{N_c k^4}{3 \pi^2}\bigg\{ \frac{2}{\sqrt{k^2+m_l^2}}\Big[ n_F(\frac{m_l^2}{k^2},T,\mu_,L,\bar L) \nonumber\\
	&   + n_F(\frac{m_l^2}{k^2},T,-\mu_,\bar L,L)\Big] + \frac{1}{\sqrt{k^2+m_s^2}}\Big[ n_F(\frac{m_s^2}{k^2},T,\mu_,L,\bar L) \nonumber\\
	&   + n_F(\frac{m_s^2}{k^2},T,-\mu_,\bar L,L)\Big]   \bigg\} \,.
\end{align}
Generally speaking, $m_{l} $ and $m_{s} $ depend on the $\rho_{1} $ and  $\rho_{2} $. It is a good approximation to set the values to their respective vacuum constituent quark masses~\cite{Fu:2018qsk}. Thus we set $m_l\sim300 $MeV and $m_s\sim430 $MeV in our calculation.

\section{The influence of parameter $\hat\gamma$ in equation (\ref{eq:bmu})}\label{app:influ}
%
\begin{figure*}[htb]
	\includegraphics[width=0.43\textwidth]{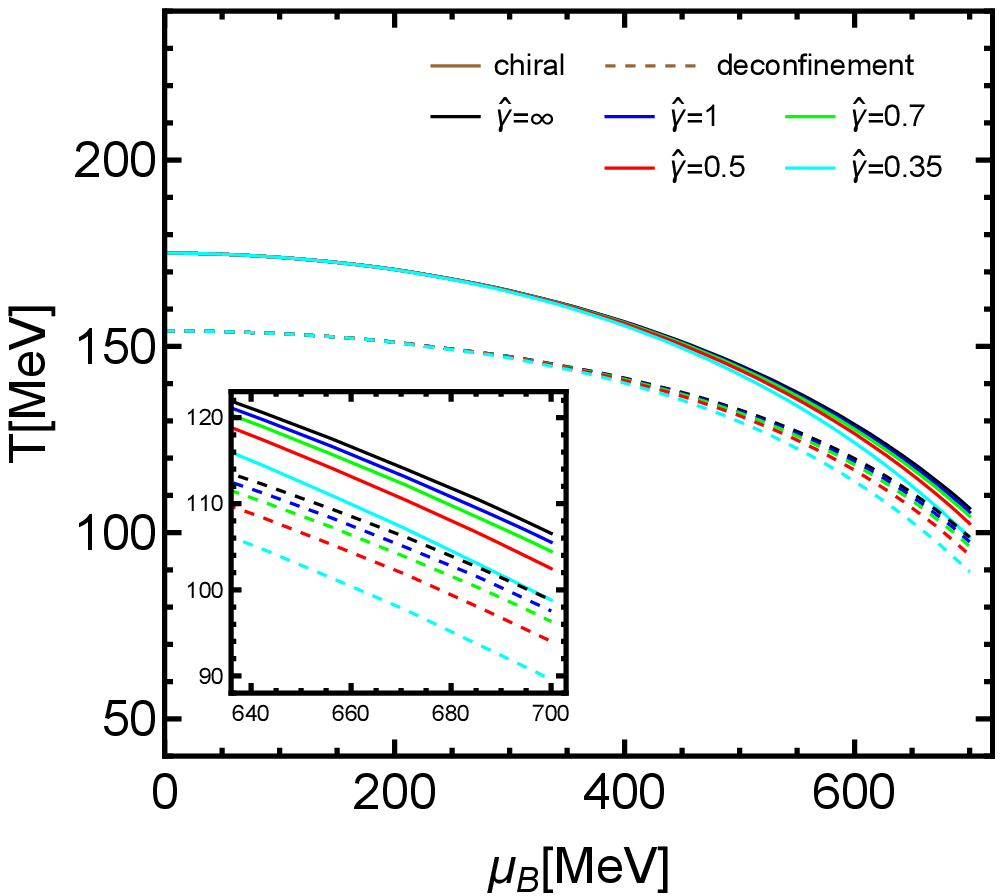} \hspace*{5mm}
	\includegraphics[width=0.43\textwidth]{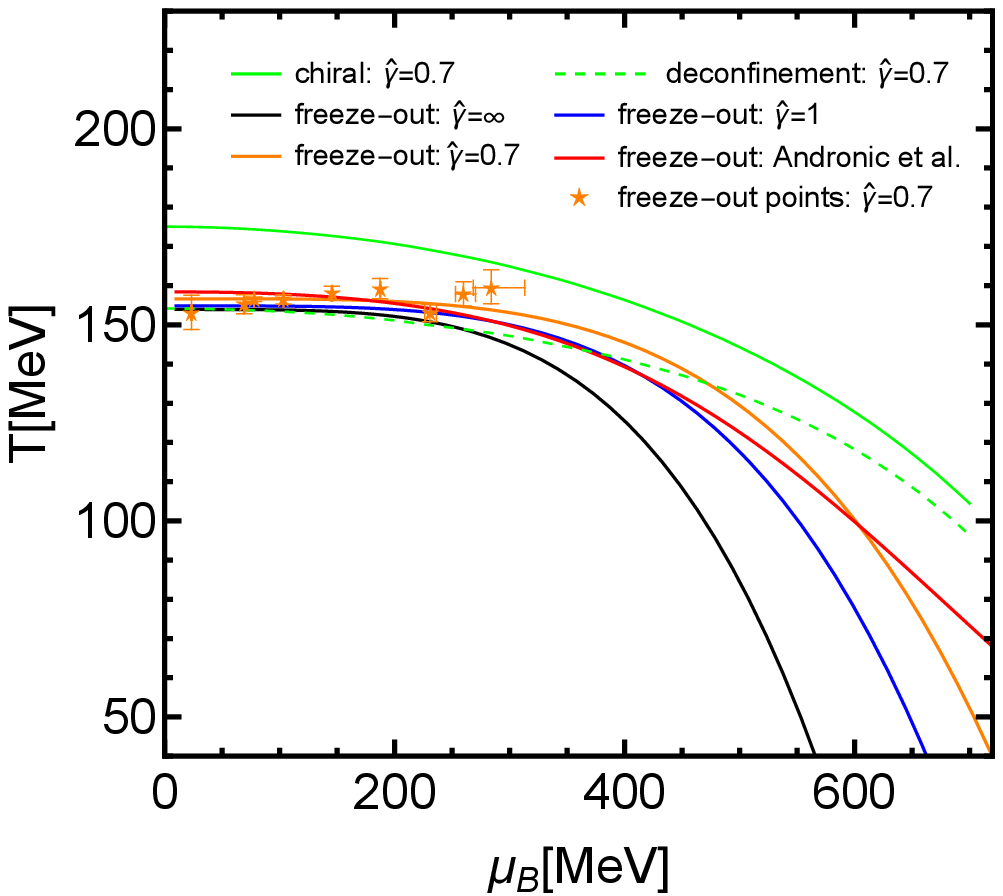}
	\caption{Left panel: The black, blue, green, red and cyan colors correspond to $\hat\gamma=\infty, 1, 0.7, 0.5$ and $0.35$, respectively. The dashed lines represent the deconfinement phase transitions while the solid lines represent the chiral phase transitions. The inlay zooms in at large $\mu_{B}$. 
		Right panel: The obtained freeze-out points and line in the case of $\hat\gamma=0.7$ (the orange stars and curve). For the convenience of comparison, we also depict the freeze-out lines of $\hat\gamma=\infty$ and $\hat\gamma=1$ in the figure (black and blue curves). Red curve is the freeze-out line from Andronic {\it et al.}~\cite{Andronic:2017pug}. The solid and dashed green lines represent the chiral and deconfinement phase transitions in the case of $\hat\gamma=0.7$, respectively.     }
	\label{fig:phasediagrams}
	\label{fig:freeze-out-07}
\end{figure*}
%

%
\begin{figure*}[htb]
	\includegraphics[width=0.42\textwidth]{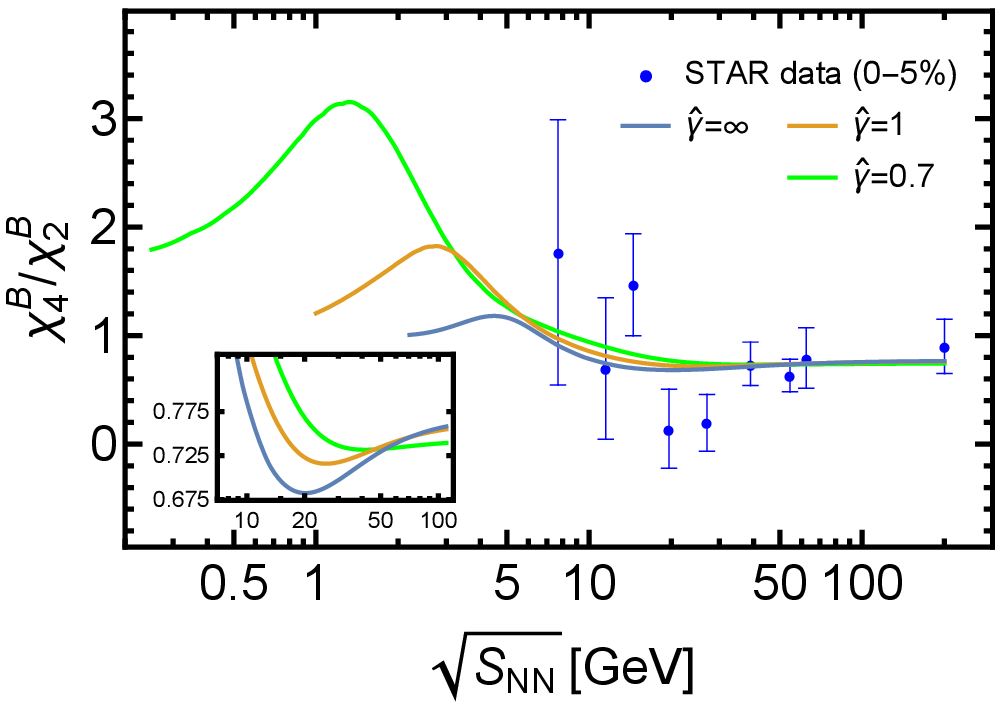} \hspace*{5mm}
	\includegraphics[width=0.45\textwidth]{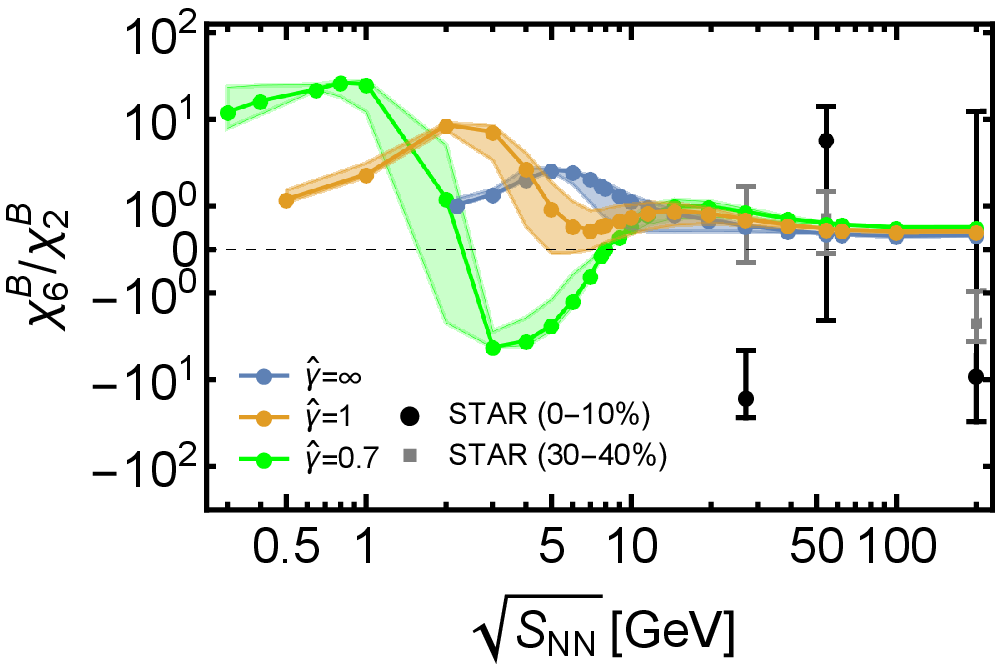}
	\caption{Left panel: Calculated collision energy dependence of $\chi_4^B/\chi_2^B$ along the freeze-out line in the case of $\hat\gamma=0.7$ (the green curve). The blue circles are the experimental values~\cite{STAR:2020tga} for $0\%-5\%$ centrality. The inlay zooms in at $\sqrt{S_{NN}}\sim 20\,$GeV.
		Right panel: Calculated collision energy dependence of $\chi_6^B/\chi_2^B$ (in logarithmic scale) along the freeze-out line in the case of $\hat\gamma=0.7$ (the green curve). The shadowed region represents our numerical uncertainties. The black circles and gray squares are the experimental values~\cite{STAR:2021rls} for $0\%-10\%$ and $30\%-40\%$ centralities, respectively.	
		For the convenience of comparison, we also depict the results of $\hat\gamma=\infty$ and $\hat\gamma=1$ in the figure.    }
	\label{fig:compareR42-07}
	\label{fig:compareR62-07}
\end{figure*}
%

%
\begin{table*}[t!]
	\renewcommand{\arraystretch}{1.3}
	\caption{Obtained freeze-out points $(\mu_B^f, T^f)$ for $\hat\gamma=0.7$ and $0.5$ cases ($\mu_B^f,T^f$ are in unit MeV and $\sqrt{S_{NN}}$ in GeV). }
	\label{table:freeze-out points plus}
	\begin{tabular}{|cc|c|c|c|c|c|c|c|c|c|}
		\hline
		\multicolumn{2}{|c|}{$\sqrt{S_{NN}}$}                                & $200$ & $62.4$ & $54.4$ & $39$ & $27$ & $19.6$ & $14.5$ & $11.5$ & $7.7$ \\ \hline
		\multicolumn{1}{|c|}{\multirow{2}{*}{$\hat\gamma=0.7$}}   & $\mu_B^f$ 	& $23.3_{-0.6}^{+0.7}$ & $69.3_{-0.8}^{+0.9}$ & $77.6_{-0.2}^{+0.2}$ & $103.3_{-0.5}^{+0.5}$ & $145.3_{-0.4}^{+0.5}$ & $187.3_{-0.0}^{+0.6}$ & $230.7_{-4.5}^{+5.7}$ & $259.9_{-7.0}^{+10.5}$ & $284.0_{-15.6}^{+29.1}$    \\ \cline{2-11}
		\multicolumn{1}{|c|}{}                                   & $T^f$     & $153.1_{-4.3}^{+4.5}$ & $155.2_{-2.3}^{+2.5}$ & $156.5_{-0.5}^{+0.6}$ & $156.5_{-1.2}^{+1.2}$ & $158.0_{-1.7}^{+1.9}$ & $159.1_{-2.4}^{+2.7}$ & $152.9_{-2.1}^{+2.3}$ & $157.8_{-3.0}^{+3.2}$ & $159.4_{-4.1}^{+4.6}$      \\ \hline
		\multicolumn{1}{|c|}{\multirow{2}{*}{$\hat\gamma=0.5$}} & $\mu_B^f$	 & $23.5_{-0.7}^{+0.8}$ & $70.2_{-1.0}^{+1.2}$ & $78.8_{-0.3}^{+0.3}$ & $105.2_{-0.7}^{+0.9}$ & $149.9_{-1.4}^{+1.7}$ & $195.6_{-1.0}^{+0.9}$ & $235.3_{-1.6}^{+2.6}$ & $271.9_{-6.1}^{+7.8}$ & $299.9_{-17.8}^{+35.7}$    \\ \cline{2-11}
		\multicolumn{1}{|c|}{}                                   & $T^f$	 & $154.4_{-4.5}^{+5.3}$ & $157.1_{-2.7}^{+3.3}$ & $158.9_{-0.7}^{+0.7}$ & $159.5_{-1.6}^{+1.8}$ & $163.7_{-2.8}^{+3.0}$ & $167.6_{-3.2}^{+2.9}$ & $161.5_{-2.9}^{+2.8}$ & $167.1_{-2.6}^{+2.8}$ & $168.0_{-4.1}^{+5.0}$     \\ \hline
	\end{tabular}
	\renewcommand{\arraystretch}{1.0}
\end{table*}

\begin{table}[htb!]
	\renewcommand{\arraystretch}{1.1}
	\caption{Parameters $T_0, a, b$ in Eq.~(\ref{eq:fitmodel-tmu}) and $c, d$ in Eq.~(\ref{eq:fitmodel-mue}) for the case of $\hat\gamma=0.7$ ($c, T_0^f$ are in unit MeV and $d$ in $\mathrm{GeV}^{-1}$). }
	\label{table:freeze-out pars plus}
	\begin{tabular}{lccccc}
		\hline\hline
		~~$T_0$~~	&	~~$c$~~	&	~~$d$~~	&	~~$T_{0}^{f}$~~	&	~~$a$~~		& $b$	\\
		\hline
		$\hat\gamma=0.7$ &	$690.76$ & $0.141$	& $156.6$	&	$4.06\times10^{-8}$	& $0.0017$\\
		\hline\hline		
	\end{tabular}
	\renewcommand{\arraystretch}{1.0}
\end{table}
%

In the main text we do the calculation and compare the results in both $\mu$-dependent and $\mu$-independent $T_{0}$ cases. 
The former case corresponds to $\hat\gamma=1$ in Eq.~(\ref{eq:bmu}), which has been used in Refs.~\cite{Fu:2018swz,Fu:2018qsk} and the obtained results have found a good agreement with those of the lattice QCD and the HRG model. The latter case corresponds to $\hat\gamma=\infty$ with no hard thermal or dense loop improvements incorporated. For completeness, we also explore the influence of other values of $\hat{\gamma}$. We choose the range of $\hat\gamma$ to be $\hat\gamma\in (0.35, \infty)$. The lower bound corresponds to an unphysical value, at which $T_0(\mu)$ in Eq.~(\ref{eq:T0mu}) already tends to be zero at $\mu\sim 300$MeV.

We illustrate the phase structures corresponding to $\hat\gamma=\infty, 1, 0.7, 0.5$ and $ 0.35$ in Fig.~\ref{fig:phasediagrams}. 
We observe that the phase structures are not sensitive to large values of $\hat{\gamma}$, 
and are mildly changed for small $\mu_{B}$, while for larger $\mu_{B}$ a smaller value of $\hat\gamma$ will result in lower transition lines in the $T-\mu_{B}$ plane.
The sensitivity of the phase structures to $\hat{\gamma}$ shown here is less than that in Ref.~\cite{Herbst:2013ail}. 
This has some relevance with the parametrization of the Polyakov-loop potential. 
The fractional polynomial dependence on $T_{0}$ is encoded in Eq.~(\ref{eq:xT}), while it is polynomial dependence in Eq.~(5) in Ref.~\cite{Herbst:2013ail}. 
Also the currently used Polyakov-loop potential has more subtle refinements compared to the one used there, such as the inclusion of the Haar measure and rescaling of the deconfinement critical temperature, see for example Eqs.~(\ref{eq:haar}) and (\ref{eq:rescalet}).

Since there is a direct contribution from $\mu$-dependent Polyakov-loop potential to the baryon number susceptibilities, the influence of $\hat\gamma$ on the freeze-out line will be more apparent than that on the phase structures. We collect the obtained freeze-out points for $\hat\gamma=0.7$ and $\hat\gamma=0.5$ cases in Table~\ref{table:freeze-out points plus} and depict the freeze-out line for $\hat\gamma=0.7$ case in Fig.~\ref{fig:freeze-out-07}. One observes that the freeze-out temperatures are overall higher when compared to the results obtained in the cases of $\hat\gamma=\infty$ and $\hat\gamma=1$, and the freeze-out line extends to larger $\mu_B$ in the low temperature region.

We do not explore the lower values of $\hat{\gamma}$. As one can see from Fig.~\ref{fig:freeze-out-07}, the freeze-out temperatures show a slowly increasing trend with the decreasing collision energy for the first sixth freeze-out points already in $\hat{\gamma}=0.7$ case. The trend is much more apparent with lower values of $\hat\gamma$, see for example the obtained freeze-out points in the case of $\hat\gamma=0.5$ in Table~\ref{table:freeze-out points plus}. The trend continues to large $\mu_B$ and thus is not physical, and we think the value $\hat\gamma=0.7$ has already come close to the unphysical one in the present settings. Therefore we refrain ourselves from lower values of $\hat{\gamma}$.

We collect the fitting parameters in Eqs.~(\ref{eq:fitmodel-tmu}) and (\ref{eq:fitmodel-mue}) in the case of $\hat\gamma=0.7$ in Table~\ref{table:freeze-out pars plus}. Then we calculate $\chi_4^B/\chi_2^B$ and $\chi_6^B/\chi_2^B$ along the freeze-out line and illustrate the $\sqrt{S_{NN}}$ dependence of $\chi_4^B/\chi_2^B$ and $\chi_6^B/\chi_2^B$ in Fig.~\ref{fig:compareR42-07}.
We observe from Fig.~\ref{fig:compareR42-07} that although the overall shapes of $\chi_4^B/\chi_2^B$ and $\chi_6^B/\chi_2^B$ are similar to those obtained in the cases of $\hat\gamma=\infty$ and $\hat\gamma=1$, the nonmonotonicity of $\chi_4^B/\chi_2^B$ shown in the inlay of the left panel of Fig.~\ref{fig:compareR42-07} is much weaker, and the minimum or maximum of $\chi_6^B/\chi_2^B$ tends to lower collision energy with larger depth or height.

	


\begin{thebibliography}{88}

	\makeatletter
	\providecommand \@ifxundefined [1]{%
		\@ifx{#1\undefined}
	}%
	\providecommand \@ifnum [1]{%
		\ifnum #1\expandafter \@firstoftwo
		\else \expandafter \@secondoftwo
		\fi
	}%
	\providecommand \@ifx [1]{%
		\ifx #1\expandafter \@firstoftwo
		\else \expandafter \@secondoftwo
		\fi
	}%
	\providecommand \natexlab [1]{#1}%
	\providecommand \enquote  [1]{``#1''}%
	\providecommand \bibnamefont  [1]{#1}%
	\providecommand \bibfnamefont [1]{#1}%
	\providecommand \citenamefont [1]{#1}%
	\providecommand \href@noop [0]{\@secondoftwo}%
	\providecommand \href [0]{\begingroup \@sanitize@url \@href}%
	\providecommand \@href[1]{\@@startlink{#1}\@@href}%
	\providecommand \@@href[1]{\endgroup#1\@@endlink}%
	\providecommand \@sanitize@url [0]{\catcode `\\12\catcode `\$12\catcode
		`\&12\catcode `\#12\catcode `\^12\catcode `\_12\catcode `\%12\relax}%
	\providecommand \@@startlink[1]{}%
	\providecommand \@@endlink[0]{}%
	\providecommand \url  [0]{\begingroup\@sanitize@url \@url }%
	\providecommand \@url [1]{\endgroup\@href {#1}{\urlprefix }}%
	\providecommand \urlprefix  [0]{URL }%
	\providecommand \Eprint [0]{\href }%
	\providecommand \doibase [0]{http://dx.doi.org/}%
	\providecommand \selectlanguage [0]{\@gobble}%
	\providecommand \bibinfo  [0]{\@secondoftwo}%
	\providecommand \bibfield  [0]{\@secondoftwo}%
	\providecommand \translation [1]{[#1]}%
	\providecommand \BibitemOpen [0]{}%
	\providecommand \bibitemStop [0]{}%
	\providecommand \bibitemNoStop [0]{.\EOS\space}%
	\providecommand \EOS [0]{\spacefactor3000\relax}%
	\providecommand \BibitemShut  [1]{\csname bibitem#1\endcsname}%
	\let\auto@bib@innerbib\@empty
%
	\bibitem [{\citenamefont {Schwarz}(2003)}]{Schwarz:2003du}%
	\BibitemOpen
	\bibfield  {author} {\bibinfo {author} {\bibfnamefont {D.~J.}\ \bibnamefont
			{Schwarz}},\ }\href {\doibase 10.1002/andp.200310010} {\bibfield  {journal}
		{\bibinfo  {journal} {Annalen Phys.}\ }\textbf {\bibinfo {volume} {12}},\
		\bibinfo {pages} {220} (\bibinfo {year} {2003})},\ \Eprint
	{http://arxiv.org/abs/0303574} {arXiv:0303574 [astro-ph]} \BibitemShut
	{NoStop}%
%
	\bibitem [{\citenamefont {Bazavov}\ and\ \citenamefont
		{Others}(2019)}]{HotQCD:2018pds}%
	\BibitemOpen
	\bibfield  {author} {\bibinfo {author} {\bibfnamefont {A.}~\bibnamefont
			{Bazavov}}\ and\ \bibinfo {author} {\bibnamefont {Others}},\ }\href {\doibase
		10.1016/j.physletb.2019.05.013} {\bibfield  {journal} {\bibinfo  {journal}
			{Phys. Lett. B}\ }\textbf {\bibinfo {volume} {795}},\ \bibinfo {pages} {15}
		(\bibinfo {year} {2019})},\ \Eprint {http://arxiv.org/abs/1812.08235}
	{arXiv:1812.08235 [hep-lat]} \BibitemShut {NoStop}%
%
	\bibitem [{\citenamefont {Ding}\ and\ \citenamefont
		{Others}(2019)}]{HotQCD:2019xnw}%
	\BibitemOpen
	\bibfield  {author} {\bibinfo {author} {\bibfnamefont {H.~T.}\ \bibnamefont
			{Ding}}\ and\ \bibinfo {author} {\bibnamefont {Others}},\ }\href {\doibase
		10.1103/PhysRevLett.123.062002} {\bibfield  {journal} {\bibinfo  {journal}
			{Phys. Rev. Lett.}\ }\textbf {\bibinfo {volume} {123}},\ \bibinfo {pages}
		{062002} (\bibinfo {year} {2019})},\ \Eprint {http://arxiv.org/abs/1903.04801}
	{arXiv:1903.04801 [hep-lat]} \BibitemShut {NoStop}%
%
	\bibitem [{\citenamefont {Muroya}\ \emph {et~al.}(2003)\citenamefont {Muroya},
		\citenamefont {Nakamura}, \citenamefont {Nonaka},\ and\ \citenamefont
		{Takaishi}}]{Muroya:2003qs}%
	\BibitemOpen
	\bibfield  {author} {\bibinfo {author} {\bibfnamefont {S.}~\bibnamefont
			{Muroya}}, \bibinfo {author} {\bibfnamefont {A.}~\bibnamefont {Nakamura}},
		\bibinfo {author} {\bibfnamefont {C.}~\bibnamefont {Nonaka}}, \ and\ \bibinfo
		{author} {\bibfnamefont {T.}~\bibnamefont {Takaishi}},\ }\href {\doibase
		10.1143/PTP.110.615} {\bibfield  {journal} {\bibinfo  {journal} {Prog. Theor.
				Phys.}\ }\textbf {\bibinfo {volume} {110}},\ \bibinfo {pages} {615} (\bibinfo
		{year} {2003})},\ \Eprint {http://arxiv.org/abs/0306031} {arXiv:0306031
		[hep-lat]} \BibitemShut {NoStop}%
%
	\bibitem [{\citenamefont {Splittorff}\ and\ \citenamefont
		{Verbaarschot}(2007)}]{Splittorff:2007ck}%
	\BibitemOpen
	\bibfield  {author} {\bibinfo {author} {\bibfnamefont {K.}~\bibnamefont
			{Splittorff}}\ and\ \bibinfo {author} {\bibfnamefont {J.~J.~M.}\ \bibnamefont
			{Verbaarschot}},\ }\href {\doibase 10.1103/PhysRevD.75.116003} {\bibfield
		{journal} {\bibinfo  {journal} {Phys. Rev. D}\ }\textbf {\bibinfo {volume}
			{75}},\ \bibinfo {pages} {116003} (\bibinfo {year} {2007})},\ \Eprint
	{http://arxiv.org/abs/0702011} {arXiv:0702011 [hep-lat]} \BibitemShut
	{NoStop}%
%
	\bibitem [{\citenamefont {Fu}(2022)}]{Fu:2022gou}%
	\BibitemOpen
	\bibfield  {author} {\bibinfo {author} {\bibfnamefont {W.-j.}\ \bibnamefont
			{Fu}},\ }\href {\doibase 10.1088/1572-9494/ac86be} {\bibfield  {journal}
		{\bibinfo  {journal} {Commun. Theor. Phys.}\ }\textbf {\bibinfo {volume}
			{74}},\ \bibinfo {pages} {097304} (\bibinfo {year} {2022})},\ \Eprint
	{http://arxiv.org/abs/2205.00468} {arXiv:2205.00468 [hep-ph]} \BibitemShut
	{NoStop}%
%
	\bibitem [{\citenamefont {Fu}\ \emph {et~al.}(2021)\citenamefont {Fu},
		\citenamefont {Luo}, \citenamefont {Pawlowski}, \citenamefont {Rennecke},
		\citenamefont {Wen},\ and\ \citenamefont {Yin}}]{Fu:2021oaw}%
	\BibitemOpen
	\bibfield  {author} {\bibinfo {author} {\bibfnamefont {W.-j.}\ \bibnamefont
			{Fu}}, \bibinfo {author} {\bibfnamefont {X.}~\bibnamefont {Luo}}, \bibinfo
		{author} {\bibfnamefont {J.~M.}\ \bibnamefont {Pawlowski}}, \bibinfo {author}
		{\bibfnamefont {F.}~\bibnamefont {Rennecke}}, \bibinfo {author}
		{\bibfnamefont {R.}~\bibnamefont {Wen}}, \ and\ \bibinfo {author}
		{\bibfnamefont {S.}~\bibnamefont {Yin}},\ }\href {\doibase
		10.1103/PhysRevD.104.094047} {\bibfield  {journal} {\bibinfo  {journal}
			{Phys. Rev. D}\ }\textbf {\bibinfo {volume} {104}},\ \bibinfo {pages} {094047}
		(\bibinfo {year} {2021})},\ \Eprint {http://arxiv.org/abs/2101.06035}
	{arXiv:2101.06035 [hep-ph]} \BibitemShut {NoStop}%
%
	\bibitem [{\citenamefont {Fu}\ \emph {et~al.}(2020{\natexlab{a}})\citenamefont
		{Fu}, \citenamefont {Pawlowski},\ and\ \citenamefont
		{Rennecke}}]{Fu:2019hdw}%
	\BibitemOpen
	\bibfield  {author} {\bibinfo {author} {\bibfnamefont {W.-j.}\ \bibnamefont
			{Fu}}, \bibinfo {author} {\bibfnamefont {J.~M.}\ \bibnamefont {Pawlowski}}, \
		and\ \bibinfo {author} {\bibfnamefont {F.}~\bibnamefont {Rennecke}},\ }\href
	{\doibase 10.1103/PhysRevD.101.054032} {\bibfield  {journal} {\bibinfo
			{journal} {Phys. Rev. D}\ }\textbf {\bibinfo {volume} {101}},\ \bibinfo
		{pages} {054032} (\bibinfo {year} {2020}{\natexlab{a}})},\ \Eprint
	{http://arxiv.org/abs/1909.02991} {arXiv:1909.02991 [hep-ph]} \BibitemShut
	{NoStop}%
%
	\bibitem [{\citenamefont {Fu}\ \emph {et~al.}(2016)\citenamefont {Fu},
		\citenamefont {Pawlowski}, \citenamefont {Rennecke},\ and\ \citenamefont
		{Schaefer}}]{Fu:2016tey}%
	\BibitemOpen
	\bibfield  {author} {\bibinfo {author} {\bibfnamefont {W.-j.}\ \bibnamefont
			{Fu}}, \bibinfo {author} {\bibfnamefont {J.~M.}\ \bibnamefont {Pawlowski}},
		\bibinfo {author} {\bibfnamefont {F.}~\bibnamefont {Rennecke}}, \ and\
		\bibinfo {author} {\bibfnamefont {B.-J.}\ \bibnamefont {Schaefer}},\ }\href
	{\doibase 10.1103/PhysRevD.94.116020} {\bibfield  {journal} {\bibinfo
			{journal} {Phys. Rev. D}\ }\textbf {\bibinfo {volume} {94}},\ \bibinfo
		{pages} {116020} (\bibinfo {year} {2016})},\ \Eprint
	{http://arxiv.org/abs/1608.04302} {arXiv:1608.04302 [hep-ph]} \BibitemShut
	{NoStop}%
%
	\bibitem [{\citenamefont {Chen}\ \emph {et~al.}(2021)\citenamefont {Chen},
		\citenamefont {Wen},\ and\ \citenamefont {Fu}}]{Chen:2021iuo}%
	\BibitemOpen
	\bibfield  {author} {\bibinfo {author} {\bibfnamefont {Y.-r.}\ \bibnamefont
			{Chen}}, \bibinfo {author} {\bibfnamefont {R.}~\bibnamefont {Wen}}, \ and\
		\bibinfo {author} {\bibfnamefont {W.-j.}\ \bibnamefont {Fu}},\ }\href
	{\doibase 10.1103/PhysRevD.104.054009} {\bibfield  {journal} {\bibinfo
			{journal} {Phys. Rev. D}\ }\textbf {\bibinfo {volume} {104}},\ \bibinfo
		{pages} {054009} (\bibinfo {year} {2021})},\ \Eprint
	{http://arxiv.org/abs/2101.08484} {arXiv:2101.08484 [hep-ph]} \BibitemShut
	{NoStop}%
%
	\bibitem [{\citenamefont {Pawlowski}\ and\ \citenamefont
		{Rennecke}(2014)}]{Pawlowski:2014zaa}%
	\BibitemOpen
	\bibfield  {author} {\bibinfo {author} {\bibfnamefont {J.~M.}\ \bibnamefont
			{Pawlowski}}\ and\ \bibinfo {author} {\bibfnamefont {F.}~\bibnamefont
			{Rennecke}},\ }\href {\doibase 10.1103/PhysRevD.90.076002} {\bibfield
		{journal} {\bibinfo  {journal} {Phys. Rev. D}\ }\textbf {\bibinfo {volume}
			{90}},\ \bibinfo {pages} {076002} (\bibinfo {year} {2014})},\ \Eprint
	{http://arxiv.org/abs/1403.1179} {arXiv:1403.1179 [hep-ph]} \BibitemShut
	{NoStop}%
%
	\bibitem [{\citenamefont {Rennecke}\ and\ \citenamefont
		{Schaefer}(2017)}]{Rennecke:2016tkm}%
	\BibitemOpen
	\bibfield  {author} {\bibinfo {author} {\bibfnamefont {F.}~\bibnamefont
			{Rennecke}}\ and\ \bibinfo {author} {\bibfnamefont {B.-J.}\ \bibnamefont
			{Schaefer}},\ }\href {\doibase 10.1103/PhysRevD.96.016009} {\bibfield
		{journal} {\bibinfo  {journal} {Phys. Rev. D}\ }\textbf {\bibinfo {volume}
			{96}},\ \bibinfo {pages} {016009} (\bibinfo {year} {2017})},\ \Eprint
	{http://arxiv.org/abs/1610.08748} {arXiv:1610.08748 [hep-ph]} \BibitemShut
	{NoStop}%
%
	\bibitem [{\citenamefont {Fu}\ \emph {et~al.}(2020{\natexlab{b}})\citenamefont
		{Fu}, \citenamefont {Pawlowski},\ and\ \citenamefont
		{Rennecke}}]{Fu:2018qsk}%
	\BibitemOpen
	\bibfield  {author} {\bibinfo {author} {\bibfnamefont {W.-j.}\ \bibnamefont
			{Fu}}, \bibinfo {author} {\bibfnamefont {J.~M.}\ \bibnamefont {Pawlowski}}, \
		and\ \bibinfo {author} {\bibfnamefont {F.}~\bibnamefont {Rennecke}},\ }\href
	{\doibase 10.21468/SciPostPhysCore.2.1.002} {\bibfield  {journal} {\bibinfo
			{journal} {SciPost Phys. Core}\ }\textbf {\bibinfo {volume} {2}},\ \bibinfo
		{pages} {2} (\bibinfo {year} {2020}{\natexlab{b}})},\ \Eprint
	{http://arxiv.org/abs/1808.00410} {arXiv:1808.00410 [hep-ph]} \BibitemShut
	{NoStop}%
%
	\bibitem [{\citenamefont {Isserstedt}\ \emph {et~al.}(2019)\citenamefont
		{Isserstedt}, \citenamefont {Buballa}, \citenamefont {Fischer},\ and\
		\citenamefont {Gunkel}}]{Isserstedt:2019pgx}%
	\BibitemOpen
	\bibfield  {author} {\bibinfo {author} {\bibfnamefont {P.}~\bibnamefont
			{Isserstedt}}, \bibinfo {author} {\bibfnamefont {M.}~\bibnamefont {Buballa}},
		\bibinfo {author} {\bibfnamefont {C.~S.}\ \bibnamefont {Fischer}}, \ and\
		\bibinfo {author} {\bibfnamefont {P.~J.}\ \bibnamefont {Gunkel}},\ }\href
	{\doibase 10.1103/PhysRevD.100.074011} {\bibfield  {journal} {\bibinfo
			{journal} {Phys. Rev. D}\ }\textbf {\bibinfo {volume} {100}},\ \bibinfo
		{pages} {074011} (\bibinfo {year} {2019})},\ \Eprint
	{http://arxiv.org/abs/1906.11644} {arXiv:1906.11644 [hep-ph]} \BibitemShut
	{NoStop}%
%
	\bibitem [{\citenamefont {Fischer}\ \emph {et~al.}(2011)\citenamefont
		{Fischer}, \citenamefont {Luecker},\ and\ \citenamefont
		{Mueller}}]{Fischer:2011mz}%
	\BibitemOpen
	\bibfield  {author} {\bibinfo {author} {\bibfnamefont {C.~S.}\ \bibnamefont
			{Fischer}}, \bibinfo {author} {\bibfnamefont {J.}~\bibnamefont {Luecker}}, \
		and\ \bibinfo {author} {\bibfnamefont {J.~A.}\ \bibnamefont {Mueller}},\
	}\href {\doibase 10.1016/j.physletb.2011.07.039} {\bibfield  {journal}
		{\bibinfo  {journal} {Phys. Lett. B}\ }\textbf {\bibinfo {volume} {702}},\
		\bibinfo {pages} {438} (\bibinfo {year} {2011})},\ \Eprint
	{http://arxiv.org/abs/1104.1564} {arXiv:1104.1564 [hep-ph]} \BibitemShut
	{NoStop}%
%
	\bibitem [{\citenamefont {Xin}\ \emph {et~al.}(2014{\natexlab{a}})\citenamefont
		{Xin}, \citenamefont {Qin},\ and\ \citenamefont {Liu}}]{Xin:2014ela}%
	\BibitemOpen
	\bibfield  {author} {\bibinfo {author} {\bibfnamefont {X.-y.}\ \bibnamefont
			{Xin}}, \bibinfo {author} {\bibfnamefont {S.-x.}\ \bibnamefont {Qin}}, \ and\
		\bibinfo {author} {\bibfnamefont {Y.-x.}\ \bibnamefont {Liu}},\ }\href
	{\doibase 10.1103/PhysRevD.90.076006} {\bibfield  {journal} {\bibinfo
			{journal} {Phys. Rev. D}\ }\textbf {\bibinfo {volume} {90}},\ \bibinfo
		{pages} {076006} (\bibinfo {year} {2014}{\natexlab{a}})},\ \Eprint
	{http://arxiv.org/abs/2109.09935} {arXiv:2109.09935 [hep-ph]} \BibitemShut
	{NoStop}%
%
	\bibitem [{\citenamefont {Gao}\ and\ \citenamefont
		{Pawlowski}(2020)}]{Gao:2020qsj}%
	\BibitemOpen
	\bibfield  {author} {\bibinfo {author} {\bibfnamefont {F.}~\bibnamefont
			{Gao}}\ and\ \bibinfo {author} {\bibfnamefont {J.~M.}\ \bibnamefont
			{Pawlowski}},\ }\href {\doibase 10.1103/PhysRevD.102.034027} {\bibfield
		{journal} {\bibinfo  {journal} {Phys. Rev. D}\ }\textbf {\bibinfo {volume}
			{102}},\ \bibinfo {pages} {034027} (\bibinfo {year} {2020})},\ \Eprint
	{http://arxiv.org/abs/2002.07500} {arXiv:2002.07500 [hep-ph]} \BibitemShut
	{NoStop}%
%
	\bibitem [{\citenamefont {Davis}(2020)}]{Davis:2020fcy}%
	\BibitemOpen
	\bibfield  {author} {\bibinfo {author} {\bibfnamefont {N.}~\bibnamefont
			{Davis}},\ }\href {\doibase 10.5506/APhysPolBSupp.13.637} {\bibfield
		{journal} {\bibinfo  {journal} {Acta Phys. Polon. Supp.}\ }\textbf {\bibinfo
			{volume} {13}},\ \bibinfo {pages} {637} (\bibinfo {year} {2020})},\ \Eprint
	{http://arxiv.org/abs/2002.06636} {arXiv:2002.06636 [nucl-ex]} \BibitemShut
	{NoStop}%
%
	\bibitem [{\citenamefont {Odyniec}(2010)}]{Odyniec_2010}%
	\BibitemOpen
	\bibfield  {author} {\bibinfo {author} {\bibfnamefont {G.}~\bibnamefont
			{Odyniec}},\ }\href {\doibase 10.1088/0954-3899/37/9/094028} {\bibfield
		{journal} {\bibinfo  {journal} {J. Phys. G}\ }\textbf {\bibinfo {volume} {37}},\ \bibinfo {pages} {94028}
		(\bibinfo {year} {2010})}\BibitemShut {NoStop}%
%
	\bibitem [{\citenamefont {Adamczyk}\ and\ \citenamefont
		{Others}(2014)}]{STAR:2013gus}%
	\BibitemOpen
	\bibfield  {author} {\bibinfo {author} {\bibfnamefont {L.}~\bibnamefont
			{Adamczyk}}\ and\ \bibinfo {author} {\bibnamefont {Others}},\ }\href
	{\doibase 10.1103/PhysRevLett.112.032302} {\bibfield  {journal} {\bibinfo
			{journal} {Phys. Rev. Lett.}\ }\textbf {\bibinfo {volume} {112}},\ \bibinfo
		{pages} {32302} (\bibinfo {year} {2014})},\ \Eprint
	{http://arxiv.org/abs/1309.5681} {arXiv:1309.5681 [nucl-ex]} \BibitemShut
	{NoStop}%
%
	\bibitem [{\citenamefont {Luo}(2015)}]{Luo:2015ewa}%
	\BibitemOpen
	\bibfield  {author} {\bibinfo {author} {\bibfnamefont {X.}~\bibnamefont
			{Luo}},\ }\href {\doibase 10.22323/1.217.0019} {\bibfield  {journal}
		{\bibinfo  {journal} {PoS}\ }\textbf {\bibinfo {volume} {CPOD2014}},\
		\bibinfo {pages} {19} (\bibinfo {year} {2015})},\ \Eprint
	{http://arxiv.org/abs/1503.02558} {arXiv:1503.02558 [nucl-ex]} \BibitemShut
	{NoStop}%
%
	\bibitem [{\citenamefont {Stephanov}\ \emph {et~al.}(1999)\citenamefont
		{Stephanov}, \citenamefont {Rajagopal},\ and\ \citenamefont
		{Shuryak}}]{Stephanov:1999zu}%
	\BibitemOpen
	\bibfield  {author} {\bibinfo {author} {\bibfnamefont {M.~A.}\ \bibnamefont
			{Stephanov}}, \bibinfo {author} {\bibfnamefont {K.}~\bibnamefont
			{Rajagopal}}, \ and\ \bibinfo {author} {\bibfnamefont {E.~V.}\ \bibnamefont
			{Shuryak}},\ }\href {\doibase 10.1103/PhysRevD.60.114028} {\bibfield
		{journal} {\bibinfo  {journal} {Phys. Rev. D}\ }\textbf {\bibinfo {volume}
			{60}},\ \bibinfo {pages} {114028} (\bibinfo {year} {1999})},\ \Eprint
	{http://arxiv.org/abs/9903292} {arXiv:9903292 [hep-ph]} \BibitemShut
	{NoStop}%
%
	\bibitem [{\citenamefont {Hatta}\ and\ \citenamefont
		{Stephanov}(2003)}]{Hatta:2003wn}%
	\BibitemOpen
	\bibfield  {author} {\bibinfo {author} {\bibfnamefont {Y.}~\bibnamefont
			{Hatta}}\ and\ \bibinfo {author} {\bibfnamefont {M.~A.}\ \bibnamefont
			{Stephanov}},\ }\href {\doibase 10.1103/PhysRevLett.91.102003} {\bibfield
		{journal} {\bibinfo  {journal} {Phys. Rev. Lett.}\ }\textbf {\bibinfo
			{volume} {91}},\ \bibinfo {pages} {102003} (\bibinfo {year} {2003})},\
	\Eprint {http://arxiv.org/abs/0302002} {arXiv:0302002 [hep-ph]} \BibitemShut
	{NoStop}%
%
	\bibitem [{\citenamefont {Stephanov}(2011)}]{Stephanov:2011pb}%
	\BibitemOpen
	\bibfield  {author} {\bibinfo {author} {\bibfnamefont {M.~A.}\ \bibnamefont
			{Stephanov}},\ }\href {\doibase 10.1103/PhysRevLett.107.052301} {\bibfield
		{journal} {\bibinfo  {journal} {Phys. Rev. Lett.}\ }\textbf {\bibinfo
			{volume} {107}},\ \bibinfo {pages} {052301} (\bibinfo {year} {2011})},\
	\Eprint {http://arxiv.org/abs/1104.1627} {arXiv:1104.1627 [hep-ph]}
	\BibitemShut {NoStop}%
%
	\bibitem [{\citenamefont {Zhou}\ and\ \citenamefont
		{Yang}(2022)}]{Zhou:2022pxl}%
	\BibitemOpen
	\bibfield  {author} {\bibinfo {author} {\bibfnamefont {X.}~\bibnamefont
			{Zhou}}\ and\ \bibinfo {author} {\bibfnamefont {J.}~\bibnamefont {Yang}}
		(\bibinfo {collaboration} {HIAF project Team}),\ }\href {\doibase
		10.1007/s43673-022-00064-1} {\bibfield  {journal} {\bibinfo  {journal} {AAPPS
				Bull.}\ }\textbf {\bibinfo {volume} {32}},\ \bibinfo {pages} {35} (\bibinfo
		{year} {2022})}\BibitemShut {NoStop}%
%
	\bibitem [{\citenamefont {Adam}\ and\ \citenamefont
		{Others}(2021)}]{STAR:2020tga}%
	\BibitemOpen
	\bibfield  {author} {\bibinfo {author} {\bibfnamefont {J.}~\bibnamefont
			{Adam}}\ and\ \bibinfo {author} {\bibnamefont {Others}},\ }\href {\doibase
		10.1103/PhysRevLett.126.092301} {\bibfield  {journal} {\bibinfo  {journal}
			{Phys. Rev. Lett.}\ }\textbf {\bibinfo {volume} {126}},\ \bibinfo {pages}
		{092301} (\bibinfo {year} {2021})},\ \Eprint {http://arxiv.org/abs/2001.02852}
	{arXiv:2001.02852 [nucl-ex]} \BibitemShut {NoStop}%
%
	\bibitem [{STA(2022{\natexlab{a}})}]{STAR:2022vlo}%
	\BibitemOpen
	\href@noop {} {\  (\bibinfo {year} {2022}{\natexlab{a}})},\ \Eprint
	{http://arxiv.org/abs/2207.09837} {arXiv:2207.09837 [nucl-ex]} \BibitemShut
	{NoStop}%
%
	\bibitem [{STA(2022{\natexlab{b}})}]{STAR:2022hbp}%
	\BibitemOpen
	\href@noop {} {\  (\bibinfo {year} {2022}{\natexlab{b}})},\ \Eprint
	{http://arxiv.org/abs/2209.08058} {arXiv:2209.08058 [nucl-ex]} \BibitemShut
	{NoStop}%
%
	\bibitem [{\citenamefont {Adam}\ and\ \citenamefont
		{Others}(2019)}]{STAR:2019ans}%
	\BibitemOpen
	\bibfield  {author} {\bibinfo {author} {\bibfnamefont {J.}~\bibnamefont
			{Adam}}\ and\ \bibinfo {author} {\bibnamefont {Others}},\ }\href {\doibase
		10.1103/PhysRevC.100.014902} {\bibfield  {journal} {\bibinfo  {journal}
			{Phys. Rev. C}\ }\textbf {\bibinfo {volume} {100}},\ \bibinfo {pages} {014902}
		(\bibinfo {year} {2019})},\ \Eprint {http://arxiv.org/abs/1903.05370}
	{arXiv:1903.05370 [nucl-ex]} \BibitemShut {NoStop}%
%
	\bibitem [{\citenamefont {Adamczyk}\ and\ \citenamefont
		{Others}(2018)}]{STAR:2017tfy}%
	\BibitemOpen
	\bibfield  {author} {\bibinfo {author} {\bibfnamefont {L.}~\bibnamefont
			{Adamczyk}}\ and\ \bibinfo {author} {\bibnamefont {Others}},\ }\href
	{\doibase 10.1016/j.physletb.2018.07.066} {\bibfield  {journal} {\bibinfo
			{journal} {Phys. Lett. B}\ }\textbf {\bibinfo {volume} {785}},\ \bibinfo
		{pages} {551} (\bibinfo {year} {2018})},\ \Eprint
	{http://arxiv.org/abs/1709.00773} {arXiv:1709.00773 [nucl-ex]} \BibitemShut
	{NoStop}%
%
	\bibitem [{\citenamefont {Lu}\ \emph {et~al.}(2022)\citenamefont {Lu},
		\citenamefont {Chen}, \citenamefont {Bai}, \citenamefont {Gao},\ and\
		\citenamefont {Liu}}]{Lu:2021ium}%
	\BibitemOpen
	\bibfield  {author} {\bibinfo {author} {\bibfnamefont {Y.}~\bibnamefont
			{Lu}}, \bibinfo {author} {\bibfnamefont {M.}~\bibnamefont {Chen}}, \bibinfo
		{author} {\bibfnamefont {Z.}~\bibnamefont {Bai}}, \bibinfo {author}
		{\bibfnamefont {F.}~\bibnamefont {Gao}}, \ and\ \bibinfo {author}
		{\bibfnamefont {Y.-x.}\ \bibnamefont {Liu}},\ }\href {\doibase
		10.1103/PhysRevD.105.034012} {\bibfield  {journal} {\bibinfo  {journal}
			{Phys. Rev. D}\ }\textbf {\bibinfo {volume} {105}},\ \bibinfo {pages} {034012}
		(\bibinfo {year} {2022})},\ \Eprint {http://arxiv.org/abs/2109.09912}
	{arXiv:2109.09912 [hep-ph]} \BibitemShut {NoStop}%
%
	\bibitem [{\citenamefont {Fu}\ and\ \citenamefont
		{Pawlowski}(2016)}]{Fu:2015amv}%
	\BibitemOpen
	\bibfield  {author} {\bibinfo {author} {\bibfnamefont {W.-j.}\ \bibnamefont
			{Fu}}\ and\ \bibinfo {author} {\bibfnamefont {J.~M.}\ \bibnamefont
			{Pawlowski}},\ }\href {\doibase 10.1103/PhysRevD.93.091501} {\bibfield
		{journal} {\bibinfo  {journal} {Phys. Rev. D}\ }\textbf {\bibinfo {volume}
			{93}},\ \bibinfo {pages} {091501} (\bibinfo {year} {2016})},\ \Eprint
	{http://arxiv.org/abs/1512.08461} {arXiv:1512.08461 [hep-ph]} \BibitemShut
	{NoStop}%
%
	\bibitem [{\citenamefont {Borsanyi}\ \emph
		{et~al.}(2014{\natexlab{a}})\citenamefont {Borsanyi}, \citenamefont {Fodor},
		\citenamefont {Katz}, \citenamefont {Krieg}, \citenamefont {Ratti},\ and\
		\citenamefont {Szabo}}]{Borsanyi:2014ewa}%
	\BibitemOpen
	\bibfield  {author} {\bibinfo {author} {\bibfnamefont {S.}~\bibnamefont
			{Borsanyi}}, \bibinfo {author} {\bibfnamefont {Z.}~\bibnamefont {Fodor}},
		\bibinfo {author} {\bibfnamefont {S.~D.}\ \bibnamefont {Katz}}, \bibinfo
		{author} {\bibfnamefont {S.}~\bibnamefont {Krieg}}, \bibinfo {author}
		{\bibfnamefont {C.}~\bibnamefont {Ratti}}, \ and\ \bibinfo {author}
		{\bibfnamefont {K.~K.}\ \bibnamefont {Szabo}},\ }\href {\doibase
		10.1103/PhysRevLett.113.052301} {\bibfield  {journal} {\bibinfo  {journal}
			{Phys. Rev. Lett.}\ }\textbf {\bibinfo {volume} {113}},\ \bibinfo {pages}
		{052301} (\bibinfo {year} {2014}{\natexlab{a}})},\ \Eprint
	{http://arxiv.org/abs/1403.4576} {arXiv:1403.4576 [hep-lat]} \BibitemShut
	{NoStop}%
%
	\bibitem [{\citenamefont {Bazavov}\ and\ \citenamefont
		{Others}(2012)}]{Bazavov:2012vg}%
	\BibitemOpen
	\bibfield  {author} {\bibinfo {author} {\bibfnamefont {A.}~\bibnamefont
			{Bazavov}}\ and\ \bibinfo {author} {\bibnamefont {Others}},\ }\href {\doibase
		10.1103/PhysRevLett.109.192302} {\bibfield  {journal} {\bibinfo  {journal}
			{Phys. Rev. Lett.}\ }\textbf {\bibinfo {volume} {109}},\ \bibinfo {pages}
		{192302} (\bibinfo {year} {2012})},\ \Eprint {http://arxiv.org/abs/1208.1220}
	{arXiv:1208.1220 [hep-lat]} \BibitemShut {NoStop}%
%
	\bibitem [{\citenamefont {Gavai}\ and\ \citenamefont
		{Gupta}(2011)}]{Gavai:2010zn}%
	\BibitemOpen
	\bibfield  {author} {\bibinfo {author} {\bibfnamefont {R.~V.}\ \bibnamefont
			{Gavai}}\ and\ \bibinfo {author} {\bibfnamefont {S.}~\bibnamefont {Gupta}},\
	}\href {\doibase 10.1016/j.physletb.2011.01.006} {\bibfield  {journal}
		{\bibinfo  {journal} {Phys. Lett. B}\ }\textbf {\bibinfo {volume} {696}},\
		\bibinfo {pages} {459} (\bibinfo {year} {2011})},\ \Eprint
	{http://arxiv.org/abs/1001.3796} {arXiv:1001.3796 [hep-lat]} \BibitemShut
	{NoStop}%
%
	\bibitem [{\citenamefont {Andronic}\ \emph {et~al.}(2006)\citenamefont
		{Andronic}, \citenamefont {Braun-Munzinger},\ and\ \citenamefont
		{Stachel}}]{Andronic:2005yp}%
	\BibitemOpen
	\bibfield  {author} {\bibinfo {author} {\bibfnamefont {A.}~\bibnamefont
			{Andronic}}, \bibinfo {author} {\bibfnamefont {P.}~\bibnamefont
			{Braun-Munzinger}}, \ and\ \bibinfo {author} {\bibfnamefont {J.}~\bibnamefont
			{Stachel}},\ }\href {\doibase 10.1016/j.nuclphysa.2006.03.012} {\bibfield
		{journal} {\bibinfo  {journal} {Nucl. Phys. A}\ }\textbf {\bibinfo {volume}
			{772}},\ \bibinfo {pages} {167} (\bibinfo {year} {2006})},\ \Eprint
	{http://arxiv.org/abs/0511071} {arXiv:0511071 [nucl-th]} \BibitemShut
	{NoStop}%
%
	\bibitem [{\citenamefont {Andronic}\ \emph {et~al.}(2018)\citenamefont
		{Andronic}, \citenamefont {Braun-Munzinger}, \citenamefont {Redlich},\ and\
		\citenamefont {Stachel}}]{Andronic:2017pug}%
	\BibitemOpen
	\bibfield  {author} {\bibinfo {author} {\bibfnamefont {A.}~\bibnamefont
			{Andronic}}, \bibinfo {author} {\bibfnamefont {P.}~\bibnamefont
			{Braun-Munzinger}}, \bibinfo {author} {\bibfnamefont {K.}~\bibnamefont
			{Redlich}}, \ and\ \bibinfo {author} {\bibfnamefont {J.}~\bibnamefont
			{Stachel}},\ }\href {\doibase 10.1038/s41586-018-0491-6} {\bibfield
		{journal} {\bibinfo  {journal} {Nature}\ }\textbf {\bibinfo {volume} {561}},\
		\bibinfo {pages} {321} (\bibinfo {year} {2018})},\ \Eprint
	{http://arxiv.org/abs/1710.09425} {arXiv:1710.09425 [nucl-th]} \BibitemShut
	{NoStop}%
%
	\bibitem [{\citenamefont {Becattini}\ \emph {et~al.}(2006)\citenamefont
		{Becattini}, \citenamefont {Manninen},\ and\ \citenamefont
		{Gazdzicki}}]{Becattini:2005xt}%
	\BibitemOpen
	\bibfield  {author} {\bibinfo {author} {\bibfnamefont {F.}~\bibnamefont
			{Becattini}}, \bibinfo {author} {\bibfnamefont {J.}~\bibnamefont {Manninen}},
		\ and\ \bibinfo {author} {\bibfnamefont {M.}~\bibnamefont {Gazdzicki}},\
	}\href {\doibase 10.1103/PhysRevC.73.044905} {\bibfield  {journal} {\bibinfo
			{journal} {Phys. Rev. C}\ }\textbf {\bibinfo {volume} {73}},\ \bibinfo
		{pages} {044905} (\bibinfo {year} {2006})},\ \Eprint
	{http://arxiv.org/abs/0511092} {arXiv:0511092 [hep-ph]} \BibitemShut
	{NoStop}%
%
	\bibitem [{\citenamefont {Alba}\ \emph {et~al.}(2014)\citenamefont {Alba},
		\citenamefont {Alberico}, \citenamefont {Bellwied}, \citenamefont {Bluhm},
		\citenamefont {{Mantovani Sarti}}, \citenamefont {Nahrgang},\ and\
		\citenamefont {Ratti}}]{Alba:2014eba}%
	\BibitemOpen
	\bibfield  {author} {\bibinfo {author} {\bibfnamefont {P.}~\bibnamefont
			{Alba}}, \bibinfo {author} {\bibfnamefont {W.}~\bibnamefont {Alberico}},
		\bibinfo {author} {\bibfnamefont {R.}~\bibnamefont {Bellwied}}, \bibinfo
		{author} {\bibfnamefont {M.}~\bibnamefont {Bluhm}}, \bibinfo {author}
		{\bibfnamefont {V.}~\bibnamefont {{Mantovani Sarti}}}, \bibinfo {author}
		{\bibfnamefont {M.}~\bibnamefont {Nahrgang}}, \ and\ \bibinfo {author}
		{\bibfnamefont {C.}~\bibnamefont {Ratti}},\ }\href {\doibase
		10.1016/j.physletb.2014.09.052} {\bibfield  {journal} {\bibinfo  {journal}
			{Phys. Lett. B}\ }\textbf {\bibinfo {volume} {738}},\ \bibinfo {pages} {305}
		(\bibinfo {year} {2014})},\ \Eprint {http://arxiv.org/abs/1403.4903}
	{arXiv:1403.4903 [hep-ph]} \BibitemShut {NoStop}%
%
	\bibitem [{\citenamefont {Karsch}\ and\ \citenamefont
		{Redlich}(2011)}]{Karsch:2010ck}%
	\BibitemOpen
	\bibfield  {author} {\bibinfo {author} {\bibfnamefont {F.}~\bibnamefont
			{Karsch}}\ and\ \bibinfo {author} {\bibfnamefont {K.}~\bibnamefont
			{Redlich}},\ }\href {\doibase 10.1016/j.physletb.2010.10.046} {\bibfield
		{journal} {\bibinfo  {journal} {Phys. Lett. B}\ }\textbf {\bibinfo {volume}
			{695}},\ \bibinfo {pages} {136} (\bibinfo {year} {2011})},\ \Eprint
	{http://arxiv.org/abs/1007.2581} {arXiv:1007.2581 [hep-ph]} \BibitemShut
	{NoStop}%
%
	\bibitem [{\citenamefont {Chen}\ \emph {et~al.}(2016)\citenamefont {Chen},
		\citenamefont {Deng}, \citenamefont {Kohyama},\ and\ \citenamefont
		{Labun}}]{Chen:2015dra}%
	\BibitemOpen
	\bibfield  {author} {\bibinfo {author} {\bibfnamefont {J.-W.}\ \bibnamefont
			{Chen}}, \bibinfo {author} {\bibfnamefont {J.}~\bibnamefont {Deng}}, \bibinfo
		{author} {\bibfnamefont {H.}~\bibnamefont {Kohyama}}, \ and\ \bibinfo
		{author} {\bibfnamefont {L.}~\bibnamefont {Labun}},\ }\href {\doibase
		10.1103/PhysRevD.93.034037} {\bibfield  {journal} {\bibinfo  {journal} {Phys.
				Rev. D}\ }\textbf {\bibinfo {volume} {93}},\ \bibinfo {pages} {034037}
		(\bibinfo {year} {2016})},\ \Eprint {http://arxiv.org/abs/1509.04968}
	{arXiv:1509.04968 [hep-ph]} \BibitemShut {NoStop}%
%
	\bibitem [{\citenamefont {Fukushima}(2011)}]{Fukushima:2010is}%
	\BibitemOpen
	\bibfield  {author} {\bibinfo {author} {\bibfnamefont {K.}~\bibnamefont
			{Fukushima}},\ }\href {\doibase 10.1016/j.physletb.2010.11.040} {\bibfield
		{journal} {\bibinfo  {journal} {Phys. Lett. B}\ }\textbf {\bibinfo {volume}
			{695}},\ \bibinfo {pages} {387} (\bibinfo {year} {2011})},\ \Eprint
	{http://arxiv.org/abs/1006.2596} {arXiv:1006.2596 [hep-ph]} \BibitemShut
	{NoStop}%
	\bibitem [{\citenamefont {Li}\ \emph {et~al.}(2019)\citenamefont {Li},
		\citenamefont {Xu}, \citenamefont {Wang},\ and\ \citenamefont
		{Huang}}]{Li:2018ygx}%
	\BibitemOpen
	\bibfield  {author} {\bibinfo {author} {\bibfnamefont {Z.}~\bibnamefont
			{Li}}, \bibinfo {author} {\bibfnamefont {K.}~\bibnamefont {Xu}}, \bibinfo
		{author} {\bibfnamefont {X.}~\bibnamefont {Wang}}, \ and\ \bibinfo {author}
		{\bibfnamefont {M.}~\bibnamefont {Huang}},\ }\href {\doibase
		10.1140/epjc/s10052-019-6703-x} {\bibfield  {journal} {\bibinfo  {journal}
			{Eur. Phys. J. C}\ }\textbf {\bibinfo {volume} {79}},\ \bibinfo {pages} {245}
		(\bibinfo {year} {2019})},\ \Eprint {http://arxiv.org/abs/1801.09215}
	{arXiv:1801.09215 [hep-ph]} \BibitemShut {NoStop}%
%
	\bibitem [{\citenamefont {Fukushima}(2004)}]{Fukushima:2003fw}%
	\BibitemOpen
	\bibfield  {author} {\bibinfo {author} {\bibfnamefont {K.}~\bibnamefont
			{Fukushima}},\ }\href {\doibase 10.1016/j.physletb.2004.04.027} {\bibfield
		{journal} {\bibinfo  {journal} {Phys. Lett. B}\ }\textbf {\bibinfo {volume}
			{591}},\ \bibinfo {pages} {277} (\bibinfo {year} {2004})},\ \Eprint
	{http://arxiv.org/abs/0310121} {arXiv:0310121 [hep-ph]} \BibitemShut
	{NoStop}%
	\bibitem [{\citenamefont {Fukushima}\ and\ \citenamefont
		{Skokov}(2017)}]{Fukushima:2017csk}%
	\BibitemOpen
	\bibfield  {author} {\bibinfo {author} {\bibfnamefont {K.}~\bibnamefont
			{Fukushima}}\ and\ \bibinfo {author} {\bibfnamefont {V.}~\bibnamefont
			{Skokov}},\ }\href {\doibase 10.1016/j.ppnp.2017.05.002} {\bibfield
		{journal} {\bibinfo  {journal} {Prog. Part. Nucl. Phys.}\ }\textbf {\bibinfo
			{volume} {96}},\ \bibinfo {pages} {154} (\bibinfo {year} {2017})},\ \Eprint
	{http://arxiv.org/abs/1705.00718} {arXiv:1705.00718 [hep-ph]} \BibitemShut
	{NoStop}%
%
	\bibitem [{\citenamefont {Schaefer}\ \emph {et~al.}(2007)\citenamefont
		{Schaefer}, \citenamefont {Pawlowski},\ and\ \citenamefont
		{Wambach}}]{Schaefer:2007pw}%
	\BibitemOpen
	\bibfield  {author} {\bibinfo {author} {\bibfnamefont {B.-J.}\ \bibnamefont
			{Schaefer}}, \bibinfo {author} {\bibfnamefont {J.~M.}\ \bibnamefont
			{Pawlowski}}, \ and\ \bibinfo {author} {\bibfnamefont {J.}~\bibnamefont
			{Wambach}},\ }\href {\doibase 10.1103/PhysRevD.76.074023} {\bibfield
		{journal} {\bibinfo  {journal} {Phys. Rev. D}\ }\textbf {\bibinfo {volume}
			{76}},\ \bibinfo {pages} {074023} (\bibinfo {year} {2007})},\ \Eprint
	{http://arxiv.org/abs/0704.3234} {arXiv:0704.3234 [hep-ph]} \BibitemShut
	{NoStop}%
%
	\bibitem [{\citenamefont {Herbst}\ \emph {et~al.}(2011)\citenamefont {Herbst},
		\citenamefont {Pawlowski},\ and\ \citenamefont {Schaefer}}]{Herbst:2010rf}%
	\BibitemOpen
	\bibfield  {author} {\bibinfo {author} {\bibfnamefont {T.~K.}\ \bibnamefont
			{Herbst}}, \bibinfo {author} {\bibfnamefont {J.~M.}\ \bibnamefont
			{Pawlowski}}, \ and\ \bibinfo {author} {\bibfnamefont {B.-J.}\ \bibnamefont
			{Schaefer}},\ }\href {\doibase 10.1016/j.physletb.2010.12.003} {\bibfield
		{journal} {\bibinfo  {journal} {Phys. Lett. B}\ }\textbf {\bibinfo {volume}
			{696}},\ \bibinfo {pages} {58} (\bibinfo {year} {2011})},\ \Eprint
	{http://arxiv.org/abs/1008.0081} {arXiv:1008.0081 [hep-ph]} \BibitemShut
	{NoStop}%
%
	\bibitem [{\citenamefont {Braun}(2012)}]{Braun:2011pp}%
	\BibitemOpen
	\bibfield  {author} {\bibinfo {author} {\bibfnamefont {J.}~\bibnamefont
			{Braun}},\ }\href {\doibase 10.1088/0954-3899/39/3/033001} {\bibfield
		{journal} {\bibinfo  {journal} {J. Phys. G}\ }\textbf {\bibinfo {volume}
			{39}},\ \bibinfo {pages} {33001} (\bibinfo {year} {2012})},\ \Eprint
	{http://arxiv.org/abs/1108.4449} {arXiv:1108.4449 [hep-ph]} \BibitemShut
	{NoStop}%
	\bibitem [{\citenamefont {Pawlowski}(2007)}]{Pawlowski:2005xe}%
	\BibitemOpen
	\bibfield  {author} {\bibinfo {author} {\bibfnamefont {J.~M.}\ \bibnamefont
			{Pawlowski}},\ }\href {\doibase 10.1016/j.aop.2007.01.007} {\bibfield
		{journal} {\bibinfo  {journal} {Ann. Phys.}\ }\textbf {\bibinfo {volume}
			{322}},\ \bibinfo {pages} {2831} (\bibinfo {year} {2007})},\ \Eprint
	{http://arxiv.org/abs/0512261} {arXiv:0512261 [hep-th]} \BibitemShut
	{NoStop}%
%
	\bibitem [{\citenamefont {Berges}\ \emph {et~al.}(2002)\citenamefont {Berges},
		\citenamefont {Tetradis},\ and\ \citenamefont {Wetterich}}]{Berges:2000ew}%
	\BibitemOpen
	\bibfield  {author} {\bibinfo {author} {\bibfnamefont {J.}~\bibnamefont
			{Berges}}, \bibinfo {author} {\bibfnamefont {N.}~\bibnamefont {Tetradis}}, \
		and\ \bibinfo {author} {\bibfnamefont {C.}~\bibnamefont {Wetterich}},\ }\href
	{\doibase 10.1016/S0370-1573(01)00098-9} {\bibfield  {journal} {\bibinfo
			{journal} {Phys. Rept.}\ }\textbf {\bibinfo {volume} {363}},\ \bibinfo
		{pages} {223} (\bibinfo {year} {2002})},\ \Eprint
	{http://arxiv.org/abs/0005122} {arXiv:0005122 [hep-ph]} \BibitemShut
	{NoStop}%
%
	\bibitem [{\citenamefont {Lo}\ \emph {et~al.}(2013)\citenamefont {Lo},
		\citenamefont {Friman}, \citenamefont {Kaczmarek}, \citenamefont {Redlich},\
		and\ \citenamefont {Sasaki}}]{Lo:2013hla}%
	\BibitemOpen
	\bibfield  {author} {\bibinfo {author} {\bibfnamefont {P.~M.}\ \bibnamefont
			{Lo}}, \bibinfo {author} {\bibfnamefont {B.}~\bibnamefont {Friman}}, \bibinfo
		{author} {\bibfnamefont {O.}~\bibnamefont {Kaczmarek}}, \bibinfo {author}
		{\bibfnamefont {K.}~\bibnamefont {Redlich}}, \ and\ \bibinfo {author}
		{\bibfnamefont {C.}~\bibnamefont {Sasaki}},\ }\href {\doibase
		10.1103/PhysRevD.88.074502} {\bibfield  {journal} {\bibinfo  {journal} {Phys.
				Rev. D}\ }\textbf {\bibinfo {volume} {88}},\ \bibinfo {pages} {074502}
		(\bibinfo {year} {2013})},\ \Eprint {http://arxiv.org/abs/1307.5958}
	{arXiv:1307.5958 [hep-lat]} \BibitemShut {NoStop}%
%
	\bibitem [{\citenamefont {Roessner}\ \emph {et~al.}(2007)\citenamefont
		{Roessner}, \citenamefont {Ratti},\ and\ \citenamefont
		{Weise}}]{Roessner:2006xn}%
	\BibitemOpen
	\bibfield  {author} {\bibinfo {author} {\bibfnamefont {S.}~\bibnamefont
			{Roessner}}, \bibinfo {author} {\bibfnamefont {C.}~\bibnamefont {Ratti}}, \
		and\ \bibinfo {author} {\bibfnamefont {W.}~\bibnamefont {Weise}},\ }\href
	{\doibase 10.1103/PhysRevD.75.034007} {\bibfield  {journal} {\bibinfo
			{journal} {Phys. Rev. D}\ }\textbf {\bibinfo {volume} {75}},\ \bibinfo
		{pages} {034007} (\bibinfo {year} {2007})},\ \Eprint
	{http://arxiv.org/abs/0609281} {arXiv:0609281 [hep-ph]} \BibitemShut
	{NoStop}%
%
	\bibitem [{\citenamefont {Haas}\ \emph {et~al.}(2013)\citenamefont {Haas},
		\citenamefont {Stiele}, \citenamefont {Braun}, \citenamefont {Pawlowski},\
		and\ \citenamefont {Schaffner-Bielich}}]{Haas:2013qwp}%
	\BibitemOpen
	\bibfield  {author} {\bibinfo {author} {\bibfnamefont {L.~M.}\ \bibnamefont
			{Haas}}, \bibinfo {author} {\bibfnamefont {R.}~\bibnamefont {Stiele}},
		\bibinfo {author} {\bibfnamefont {J.}~\bibnamefont {Braun}}, \bibinfo
		{author} {\bibfnamefont {J.~M.}\ \bibnamefont {Pawlowski}}, \ and\ \bibinfo
		{author} {\bibfnamefont {J.}~\bibnamefont {Schaffner-Bielich}},\ }\href
	{\doibase 10.1103/PhysRevD.87.076004} {\bibfield  {journal} {\bibinfo
			{journal} {Phys. Rev. D}\ }\textbf {\bibinfo {volume} {87}},\ \bibinfo
		{pages} {076004} (\bibinfo {year} {2013})},\ \Eprint
	{http://arxiv.org/abs/1302.1993} {arXiv:1302.1993 [hep-ph]} \BibitemShut
	{NoStop}%
%
	\bibitem [{\citenamefont {Herbst}\ \emph {et~al.}(2013)\citenamefont {Herbst},
		\citenamefont {Pawlowski},\ and\ \citenamefont {Schaefer}}]{Herbst:2013ail}%
	\BibitemOpen
	\bibfield  {author} {\bibinfo {author} {\bibfnamefont {T.~K.}\ \bibnamefont
			{Herbst}}, \bibinfo {author} {\bibfnamefont {J.~M.}\ \bibnamefont
			{Pawlowski}}, \ and\ \bibinfo {author} {\bibfnamefont {B.-J.}\ \bibnamefont
			{Schaefer}},\ }\href {\doibase 10.1103/PhysRevD.88.014007} {\bibfield
		{journal} {\bibinfo  {journal} {Phys. Rev. D}\ }\textbf {\bibinfo {volume}
			{88}},\ \bibinfo {pages} {014007} (\bibinfo {year} {2013})},\ \Eprint
	{http://arxiv.org/abs/1302.1426} {arXiv:1302.1426 [hep-ph]} \BibitemShut
	{NoStop}%
%
	\bibitem [{\citenamefont {Herbst}\ \emph {et~al.}(2014)\citenamefont {Herbst},
		\citenamefont {Mitter}, \citenamefont {Pawlowski}, \citenamefont {Schaefer},\
		and\ \citenamefont {Stiele}}]{Herbst:2013ufa}%
	\BibitemOpen
	\bibfield  {author} {\bibinfo {author} {\bibfnamefont {T.~K.}\ \bibnamefont
			{Herbst}}, \bibinfo {author} {\bibfnamefont {M.}~\bibnamefont {Mitter}},
		\bibinfo {author} {\bibfnamefont {J.~M.}\ \bibnamefont {Pawlowski}}, \bibinfo
		{author} {\bibfnamefont {B.-J.}\ \bibnamefont {Schaefer}}, \ and\ \bibinfo
		{author} {\bibfnamefont {R.}~\bibnamefont {Stiele}},\ }\href {\doibase
		10.1016/j.physletb.2014.02.045} {\bibfield  {journal} {\bibinfo  {journal}
			{Phys. Lett. B}\ }\textbf {\bibinfo {volume} {731}},\ \bibinfo {pages} {248}
		(\bibinfo {year} {2014})},\ \Eprint {http://arxiv.org/abs/1308.3621}
	{arXiv:1308.3621 [hep-ph]} \BibitemShut {NoStop}%
%
	\bibitem [{\citenamefont {Stiele}\ and\ \citenamefont
		{Schaffner-Bielich}(2016)}]{Stiele:2016cfs}%
	\BibitemOpen
	\bibfield  {author} {\bibinfo {author} {\bibfnamefont {R.}~\bibnamefont
			{Stiele}}\ and\ \bibinfo {author} {\bibfnamefont {J.}~\bibnamefont
			{Schaffner-Bielich}},\ }\href {\doibase 10.1103/PhysRevD.93.094014}
	{\bibfield  {journal} {\bibinfo  {journal} {Phys. Rev. D}\ }\textbf {\bibinfo
			{volume} {93}},\ \bibinfo {pages} {094014} (\bibinfo {year} {2016})},\ \Eprint
	{http://arxiv.org/abs/1601.05731} {arXiv:1601.05731 [hep-ph]} \BibitemShut
	{NoStop}%
%
	\bibitem [{\citenamefont {Fu}\ \emph {et~al.}(2019)\citenamefont {Fu},
		\citenamefont {Pawlowski},\ and\ \citenamefont {Rennecke}}]{Fu:2018swz}%
	\BibitemOpen
	\bibfield  {author} {\bibinfo {author} {\bibfnamefont {W.-j.}\ \bibnamefont
			{Fu}}, \bibinfo {author} {\bibfnamefont {J.~M.}\ \bibnamefont {Pawlowski}}, \
		and\ \bibinfo {author} {\bibfnamefont {F.}~\bibnamefont {Rennecke}},\ }\href
	{\doibase 10.1103/PhysRevD.100.111501} {\bibfield  {journal} {\bibinfo
			{journal} {Phys. Rev. D}\ }\textbf {\bibinfo {volume} {100}},\ \bibinfo
		{pages} {111501} (\bibinfo {year} {2019})},\ \Eprint
	{http://arxiv.org/abs/1809.01594} {arXiv:1809.01594 [hep-ph]} \BibitemShut
	{NoStop}%
%
	\bibitem [{\citenamefont {Friman}\ \emph {et~al.}(2011)\citenamefont {Friman},
		\citenamefont {Karsch}, \citenamefont {Redlich},\ and\ \citenamefont
		{Skokov}}]{Friman:2011pf}%
	\BibitemOpen
	\bibfield  {author} {\bibinfo {author} {\bibfnamefont {B.}~\bibnamefont
			{Friman}}, \bibinfo {author} {\bibfnamefont {F.}~\bibnamefont {Karsch}},
		\bibinfo {author} {\bibfnamefont {K.}~\bibnamefont {Redlich}}, \ and\
		\bibinfo {author} {\bibfnamefont {V.}~\bibnamefont {Skokov}},\ }\href
	{\doibase 10.1140/epjc/s10052-011-1694-2} {\bibfield  {journal} {\bibinfo
			{journal} {Eur. Phys. J. C}\ }\textbf {\bibinfo {volume} {71}},\ \bibinfo
		{pages} {1694} (\bibinfo {year} {2011})},\ \Eprint
	{http://arxiv.org/abs/1103.3511} {arXiv:1103.3511 [hep-ph]} \BibitemShut
	{NoStop}%
%
	\bibitem [{\citenamefont {Almasi}\ \emph
		{et~al.}(2017{\natexlab{a}})\citenamefont {Almasi}, \citenamefont {Friman},\
		and\ \citenamefont {Redlich}}]{Almasi:2017bhq}%
	\BibitemOpen
	\bibfield  {author} {\bibinfo {author} {\bibfnamefont {G.~A.}\ \bibnamefont
			{Almasi}}, \bibinfo {author} {\bibfnamefont {B.}~\bibnamefont {Friman}}, \
		and\ \bibinfo {author} {\bibfnamefont {K.}~\bibnamefont {Redlich}},\ }\href
	{\doibase 10.1103/PhysRevD.96.014027} {\bibfield  {journal} {\bibinfo
			{journal} {Phys. Rev. D}\ }\textbf {\bibinfo {volume} {96}},\ \bibinfo
		{pages} {014027} (\bibinfo {year} {2017}{\natexlab{a}})},\ \Eprint
	{http://arxiv.org/abs/1703.05947} {arXiv:1703.05947 [hep-ph]} \BibitemShut
	{NoStop}%
%
	\bibitem [{\citenamefont {Braun}\ \emph {et~al.}(2016)\citenamefont {Braun},
		\citenamefont {Fister}, \citenamefont {Pawlowski},\ and\ \citenamefont
		{Rennecke}}]{Braun:2014ata}%
	\BibitemOpen
	\bibfield  {author} {\bibinfo {author} {\bibfnamefont {J.}~\bibnamefont
			{Braun}}, \bibinfo {author} {\bibfnamefont {L.}~\bibnamefont {Fister}},
		\bibinfo {author} {\bibfnamefont {J.~M.}\ \bibnamefont {Pawlowski}}, \ and\
		\bibinfo {author} {\bibfnamefont {F.}~\bibnamefont {Rennecke}},\ }\href
	{\doibase 10.1103/PhysRevD.94.034016} {\bibfield  {journal} {\bibinfo
			{journal} {Phys. Rev. D}\ }\textbf {\bibinfo {volume} {94}},\ \bibinfo
		{pages} {034016} (\bibinfo {year} {2016})},\ \Eprint
	{http://arxiv.org/abs/1412.1045} {arXiv:1412.1045 [hep-ph]} \BibitemShut
	{NoStop}%
%
	\bibitem [{\citenamefont {Cyrol}\ \emph {et~al.}(2018)\citenamefont {Cyrol},
		\citenamefont {Mitter}, \citenamefont {Pawlowski},\ and\ \citenamefont
		{Strodthoff}}]{Cyrol:2017qkl}%
	\BibitemOpen
	\bibfield  {author} {\bibinfo {author} {\bibfnamefont {A.~K.}\ \bibnamefont
			{Cyrol}}, \bibinfo {author} {\bibfnamefont {M.}~\bibnamefont {Mitter}},
		\bibinfo {author} {\bibfnamefont {J.~M.}\ \bibnamefont {Pawlowski}}, \ and\
		\bibinfo {author} {\bibfnamefont {N.}~\bibnamefont {Strodthoff}},\ }\href
	{\doibase 10.1103/PhysRevD.97.054015} {\bibfield  {journal} {\bibinfo
			{journal} {Phys. Rev. D}\ }\textbf {\bibinfo {volume} {97}},\ \bibinfo
		{pages} {054015} (\bibinfo {year} {2018})},\ \Eprint
	{http://arxiv.org/abs/1708.03482} {arXiv:1708.03482 [hep-ph]} \BibitemShut
	{NoStop}%
%
	\bibitem [{\citenamefont {Mitter}\ \emph {et~al.}(2015)\citenamefont {Mitter},
		\citenamefont {Pawlowski},\ and\ \citenamefont
		{Strodthoff}}]{Mitter:2014wpa}%
	\BibitemOpen
	\bibfield  {author} {\bibinfo {author} {\bibfnamefont {M.}~\bibnamefont
			{Mitter}}, \bibinfo {author} {\bibfnamefont {J.~M.}\ \bibnamefont
			{Pawlowski}}, \ and\ \bibinfo {author} {\bibfnamefont {N.}~\bibnamefont
			{Strodthoff}},\ }\href {\doibase 10.1103/PhysRevD.91.054035} {\bibfield
		{journal} {\bibinfo  {journal} {Phys. Rev. D}\ }\textbf {\bibinfo {volume}
			{91}},\ \bibinfo {pages} {054035} (\bibinfo {year} {2015})},\ \Eprint
	{http://arxiv.org/abs/1411.7978} {arXiv:1411.7978 [hep-ph]} \BibitemShut
	{NoStop}%
%
	\bibitem [{\citenamefont {Mitter}\ and\ \citenamefont
		{Schaefer}(2014)}]{Mitter:2013fxa}%
	\BibitemOpen
	\bibfield  {author} {\bibinfo {author} {\bibfnamefont {M.}~\bibnamefont
			{Mitter}}\ and\ \bibinfo {author} {\bibfnamefont {B.-J.}\ \bibnamefont
			{Schaefer}},\ }\href {\doibase 10.1103/PhysRevD.89.054027} {\bibfield
		{journal} {\bibinfo  {journal} {Phys. Rev. D}\ }\textbf {\bibinfo {volume}
			{89}},\ \bibinfo {pages} {054027} (\bibinfo {year} {2014})},\ \Eprint
	{http://arxiv.org/abs/1308.3176} {arXiv:1308.3176 [hep-ph]} \BibitemShut
	{NoStop}%
%
	\bibitem [{\citenamefont {Glazek}(2008)}]{Glazek:2008mj}%
	\BibitemOpen
	\bibfield  {author} {\bibinfo {author} {\bibfnamefont {S.~D.}\ \bibnamefont
			{Glazek}},\ }\href {\doibase 10.22323/1.061.0004} {\bibfield  {journal}
		{\bibinfo  {journal} {PoS}\ }\textbf {\bibinfo {volume} {LC2008}},\ \bibinfo
		{pages} {4} (\bibinfo {year} {2008})},\ \Eprint
	{http://arxiv.org/abs/0810.4467} {arXiv:0810.4467 [hep-th]} \BibitemShut
	{NoStop}%
%
	\bibitem [{\citenamefont {Xin}\ \emph {et~al.}(2014{\natexlab{b}})\citenamefont
		{Xin}, \citenamefont {Qin},\ and\ \citenamefont {Liu}}]{Xin:2014dia}%
	\BibitemOpen
	\bibfield  {author} {\bibinfo {author} {\bibfnamefont {X.-y.}\ \bibnamefont
			{Xin}}, \bibinfo {author} {\bibfnamefont {S.-x.}\ \bibnamefont {Qin}}, \ and\
		\bibinfo {author} {\bibfnamefont {Y.-x.}\ \bibnamefont {Liu}},\ }\href
	{\doibase 10.1103/PhysRevD.89.094012} {\bibfield  {journal} {\bibinfo
			{journal} {Phys. Rev. D}\ }\textbf {\bibinfo {volume} {89}},\ \bibinfo
		{pages} {094012} (\bibinfo {year} {2014}{\natexlab{b}})}\BibitemShut {NoStop}%
%
	\bibitem [{\citenamefont {Scavenius}\ \emph {et~al.}(2002)\citenamefont
		{Scavenius}, \citenamefont {Dumitru},\ and\ \citenamefont
		{Lenaghan}}]{Scavenius:2002ru}%
	\BibitemOpen
	\bibfield  {author} {\bibinfo {author} {\bibfnamefont {O.}~\bibnamefont
			{Scavenius}}, \bibinfo {author} {\bibfnamefont {A.}~\bibnamefont {Dumitru}},
		\ and\ \bibinfo {author} {\bibfnamefont {J.~T.}\ \bibnamefont {Lenaghan}},\
	}\href {\doibase 10.1103/PhysRevC.66.034903} {\bibfield  {journal} {\bibinfo
			{journal} {Phys. Rev. C}\ }\textbf {\bibinfo {volume} {66}},\ \bibinfo
		{pages} {034903} (\bibinfo {year} {2002})},\ \Eprint
	{http://arxiv.org/abs/0201079} {arXiv:0201079 [hep-ph]} \BibitemShut
	{NoStop}%
%
	\bibitem [{\citenamefont {Skokov}\ \emph {et~al.}(2010)\citenamefont {Skokov},
		\citenamefont {Stokic}, \citenamefont {Friman},\ and\ \citenamefont
		{Redlich}}]{Skokov:2010wb}%
	\BibitemOpen
	\bibfield  {author} {\bibinfo {author} {\bibfnamefont {V.}~\bibnamefont
			{Skokov}}, \bibinfo {author} {\bibfnamefont {B.}~\bibnamefont {Stokic}},
		\bibinfo {author} {\bibfnamefont {B.}~\bibnamefont {Friman}}, \ and\ \bibinfo
		{author} {\bibfnamefont {K.}~\bibnamefont {Redlich}},\ }\href {\doibase
		10.1103/PhysRevC.82.015206} {\bibfield  {journal} {\bibinfo  {journal} {Phys.
				Rev. C}\ }\textbf {\bibinfo {volume} {82}},\ \bibinfo {pages} {015206}
		(\bibinfo {year} {2010})},\ \Eprint {http://arxiv.org/abs/1004.2665}
	{arXiv:1004.2665 [hep-ph]} \BibitemShut {NoStop}%
%
	\bibitem [{\citenamefont {Kobayashi}\ and\ \citenamefont
		{Maskawa}(1970)}]{Kobayashi:1970ji}%
	\BibitemOpen
	\bibfield  {author} {\bibinfo {author} {\bibfnamefont {M.}~\bibnamefont
			{Kobayashi}}\ and\ \bibinfo {author} {\bibfnamefont {T.}~\bibnamefont
			{Maskawa}},\ }\href {\doibase 10.1143/PTP.44.1422} {\bibfield  {journal}
		{\bibinfo  {journal} {Prog. Theor. Phys.}\ }\textbf {\bibinfo {volume}
			{44}},\ \bibinfo {pages} {1422} (\bibinfo {year} {1970})}\BibitemShut
	{NoStop}%
%
	\bibitem [{\citenamefont {{'t Hooft}}(1976)}]{tHooft:1976rip}%
	\BibitemOpen
	\bibfield  {author} {\bibinfo {author} {\bibfnamefont {G.}~\bibnamefont {{'t
					Hooft}}},\ }\href {\doibase 10.1103/PhysRevLett.37.8} {\bibfield  {journal}
		{\bibinfo  {journal} {Phys. Rev. Lett.}\ }\textbf {\bibinfo {volume} {37}},\
		\bibinfo {pages} {8} (\bibinfo {year} {1976})}\BibitemShut {NoStop}%
%
	\bibitem [{\citenamefont {Wetterich}(2001)}]{Wetterich:2001kra}%
	\BibitemOpen
	\bibfield  {author} {\bibinfo {author} {\bibfnamefont {C.}~\bibnamefont
			{Wetterich}},\ }\href {\doibase 10.1142/S0217751X01004591} {\bibfield
		{journal} {\bibinfo  {journal} {Int. J. Mod. Phys. A}\ }\textbf {\bibinfo
			{volume} {16}},\ \bibinfo {pages} {1951} (\bibinfo {year} {2001})},\ \Eprint
	{http://arxiv.org/abs/0101178} {arXiv:0101178 [hep-ph]} \BibitemShut
	{NoStop}%
%
	\bibitem [{\citenamefont {Litim}(2001)}]{Litim:2001up}%
	\BibitemOpen
	\bibfield  {author} {\bibinfo {author} {\bibfnamefont {D.~F.}\ \bibnamefont
			{Litim}},\ }\href {\doibase 10.1103/PhysRevD.64.105007} {\bibfield  {journal}
		{\bibinfo  {journal} {Phys. Rev. D}\ }\textbf {\bibinfo {volume} {64}},\
		\bibinfo {pages} {105007} (\bibinfo {year} {2001})},\ \Eprint
	{http://arxiv.org/abs/0103195} {arXiv:0103195 [hep-th]} \BibitemShut
	{NoStop}%
%
	\bibitem [{\citenamefont {Wen}\ \emph {et~al.}(2019)\citenamefont {Wen},
		\citenamefont {Huang},\ and\ \citenamefont {Fu}}]{Wen:2018nkn}%
	\BibitemOpen
	\bibfield  {author} {\bibinfo {author} {\bibfnamefont {R.}~\bibnamefont
			{Wen}}, \bibinfo {author} {\bibfnamefont {C.}~\bibnamefont {Huang}}, \ and\
		\bibinfo {author} {\bibfnamefont {W.-J.}\ \bibnamefont {Fu}},\ }\href
	{\doibase 10.1103/PhysRevD.99.094019} {\bibfield  {journal} {\bibinfo
			{journal} {Phys. Rev. D}\ }\textbf {\bibinfo {volume} {99}},\ \bibinfo
		{pages} {094019} (\bibinfo {year} {2019})},\ \Eprint
	{http://arxiv.org/abs/1809.04233} {arXiv:1809.04233 [hep-ph]} \BibitemShut
	{NoStop}%
%
	\bibitem [{\citenamefont {Kamikado}\ and\ \citenamefont
		{Kanazawa}(2015)}]{Kamikado:2014bua}%
	\BibitemOpen
	\bibfield  {author} {\bibinfo {author} {\bibfnamefont {K.}~\bibnamefont
			{Kamikado}}\ and\ \bibinfo {author} {\bibfnamefont {T.}~\bibnamefont
			{Kanazawa}},\ }\href {\doibase 10.1007/JHEP01(2015)129} {\bibfield  {journal}
		{\bibinfo  {journal} {JHEP}\ }\textbf {\bibinfo {volume} {01}},\ \bibinfo
		{pages} {129} (\bibinfo {year} {2015})},\ \Eprint
	{http://arxiv.org/abs/1410.6253} {arXiv:1410.6253 [hep-ph]} \BibitemShut
	{NoStop}%
%
	\bibitem [{\citenamefont {Yin}\ \emph {et~al.}(2019)\citenamefont {Yin},
		\citenamefont {Wen},\ and\ \citenamefont {Fu}}]{Yin:2019ebz}%
	\BibitemOpen
	\bibfield  {author} {\bibinfo {author} {\bibfnamefont {S.}~\bibnamefont
			{Yin}}, \bibinfo {author} {\bibfnamefont {R.}~\bibnamefont {Wen}}, \ and\
		\bibinfo {author} {\bibfnamefont {W.-j.}\ \bibnamefont {Fu}},\ }\href
	{\doibase 10.1103/PhysRevD.100.094029} {\bibfield  {journal} {\bibinfo
			{journal} {Phys. Rev. D}\ }\textbf {\bibinfo {volume} {100}},\ \bibinfo
		{pages} {094029} (\bibinfo {year} {2019})},\ \Eprint
	{http://arxiv.org/abs/1907.10262} {arXiv:1907.10262 [hep-ph]} \BibitemShut
	{NoStop}%
%
	\bibitem [{\citenamefont {Resch}\ \emph {et~al.}(2019)\citenamefont {Resch},
		\citenamefont {Rennecke},\ and\ \citenamefont {Schaefer}}]{Resch:2017vjs}%
	\BibitemOpen
	\bibfield  {author} {\bibinfo {author} {\bibfnamefont {S.}~\bibnamefont
			{Resch}}, \bibinfo {author} {\bibfnamefont {F.}~\bibnamefont {Rennecke}}, \
		and\ \bibinfo {author} {\bibfnamefont {B.-J.}\ \bibnamefont {Schaefer}},\
	}\href {\doibase 10.1103/PhysRevD.99.076005} {\bibfield  {journal} {\bibinfo
			{journal} {Phys. Rev. D}\ }\textbf {\bibinfo {volume} {99}},\ \bibinfo
		{pages} {076005} (\bibinfo {year} {2019})},\ \Eprint
	{http://arxiv.org/abs/1712.07961} {arXiv:1712.07961 [hep-ph]} \BibitemShut
	{NoStop}%
%
	\bibitem [{\citenamefont {Lenaghan}\ \emph {et~al.}(2000)\citenamefont
		{Lenaghan}, \citenamefont {Rischke},\ and\ \citenamefont
		{Schaffner-Bielich}}]{Lenaghan:2000ey}%
	\BibitemOpen
	\bibfield  {author} {\bibinfo {author} {\bibfnamefont {J.~T.}\ \bibnamefont
			{Lenaghan}}, \bibinfo {author} {\bibfnamefont {D.~H.}\ \bibnamefont
			{Rischke}}, \ and\ \bibinfo {author} {\bibfnamefont {J.}~\bibnamefont
			{Schaffner-Bielich}},\ }\href {\doibase 10.1103/PhysRevD.62.085008}
	{\bibfield  {journal} {\bibinfo  {journal} {Phys. Rev. D}\ }\textbf {\bibinfo
			{volume} {62}},\ \bibinfo {pages} {085008} (\bibinfo {year} {2000})},\ \Eprint
	{http://arxiv.org/abs/0004006} {arXiv:0004006 [nucl-th]} \BibitemShut
	{NoStop}%
%
	\bibitem [{\citenamefont {Borsanyi}\ \emph
		{et~al.}(2014{\natexlab{b}})\citenamefont {Borsanyi}, \citenamefont {Fodor},
		\citenamefont {Hoelbling}, \citenamefont {Katz}, \citenamefont {Krieg},\ and\
		\citenamefont {Szabo}}]{Borsanyi:2013bia}%
	\BibitemOpen
	\bibfield  {author} {\bibinfo {author} {\bibfnamefont {S.}~\bibnamefont
			{Borsanyi}}, \bibinfo {author} {\bibfnamefont {Z.}~\bibnamefont {Fodor}},
		\bibinfo {author} {\bibfnamefont {C.}~\bibnamefont {Hoelbling}}, \bibinfo
		{author} {\bibfnamefont {S.~D.}\ \bibnamefont {Katz}}, \bibinfo {author}
		{\bibfnamefont {S.}~\bibnamefont {Krieg}}, \ and\ \bibinfo {author}
		{\bibfnamefont {K.~K.}\ \bibnamefont {Szabo}},\ }\href {\doibase
		10.1016/j.physletb.2014.01.007} {\bibfield  {journal} {\bibinfo  {journal}
			{Phys. Lett. B}\ }\textbf {\bibinfo {volume} {730}},\ \bibinfo {pages} {99}
		(\bibinfo {year} {2014}{\natexlab{b}})},\ \Eprint
	{http://arxiv.org/abs/1309.5258} {arXiv:1309.5258 [hep-lat]} \BibitemShut
	{NoStop}%
%
	\bibitem [{\citenamefont {Karsch}(2012)}]{Karsch:2012wm}%
	\BibitemOpen
	\bibfield  {author} {\bibinfo {author} {\bibfnamefont {F.}~\bibnamefont
			{Karsch}},\ }\href {\doibase 10.2478/s11534-012-0074-3} {\bibfield  {journal}
		{\bibinfo  {journal} {Central Eur. J. Phys.}\ }\textbf {\bibinfo {volume}
			{10}},\ \bibinfo {pages} {1234} (\bibinfo {year} {2012})},\ \Eprint
	{http://arxiv.org/abs/1202.4173} {arXiv:1202.4173 [hep-lat]} \BibitemShut
	{NoStop}%
%
	\bibitem [{\citenamefont {Braun-Munzinger}\ \emph {et~al.}(2021)\citenamefont
		{Braun-Munzinger}, \citenamefont {Friman}, \citenamefont {Redlich},
		\citenamefont {Rustamov},\ and\ \citenamefont
		{Stachel}}]{Braun-Munzinger:2020jbk}%
	\BibitemOpen
	\bibfield  {author} {\bibinfo {author} {\bibfnamefont {P.}~\bibnamefont
			{Braun-Munzinger}}, \bibinfo {author} {\bibfnamefont {B.}~\bibnamefont
			{Friman}}, \bibinfo {author} {\bibfnamefont {K.}~\bibnamefont {Redlich}},
		\bibinfo {author} {\bibfnamefont {A.}~\bibnamefont {Rustamov}}, \ and\
		\bibinfo {author} {\bibfnamefont {J.}~\bibnamefont {Stachel}},\ }\href
	{\doibase 10.1016/j.nuclphysa.2021.122141} {\bibfield  {journal} {\bibinfo
			{journal} {Nucl. Phys. A}\ }\textbf {\bibinfo {volume} {1008}},\ \bibinfo
		{pages} {122141} (\bibinfo {year} {2021})},\ \Eprint
	{http://arxiv.org/abs/2007.02463} {arXiv:2007.02463 [nucl-th]} \BibitemShut
	{NoStop}%
%
	\bibitem [{\citenamefont {Bazavov}\ and\ \citenamefont
		{Others}(2016)}]{Bazavov:2015zja}%
	\BibitemOpen
	\bibfield  {author} {\bibinfo {author} {\bibfnamefont {A.}~\bibnamefont
			{Bazavov}}\ and\ \bibinfo {author} {\bibnamefont {Others}},\ }\href {\doibase
		10.1103/PhysRevD.93.014512} {\bibfield  {journal} {\bibinfo  {journal} {Phys.
				Rev. D}\ }\textbf {\bibinfo {volume} {93}},\ \bibinfo {pages} {014512}
		(\bibinfo {year} {2016})},\ \Eprint {http://arxiv.org/abs/1509.05786}
	{arXiv:1509.05786 [hep-lat]} \BibitemShut {NoStop}%
%
	\bibitem [{\citenamefont {Abdallah}\ and\ \citenamefont
		{Others}(2021)}]{STAR:2021rls}%
	\BibitemOpen
	\bibfield  {author} {\bibinfo {author} {\bibfnamefont {M.}~\bibnamefont
			{Abdallah}}\ and\ \bibinfo {author} {\bibnamefont {Others}},\ }\href
	{\doibase 10.1103/PhysRevLett.127.262301} {\bibfield  {journal} {\bibinfo
			{journal} {Phys. Rev. Lett.}\ }\textbf {\bibinfo {volume} {127}},\ \bibinfo
		{pages} {262301} (\bibinfo {year} {2021})},\ \Eprint
	{http://arxiv.org/abs/2105.14698} {arXiv:2105.14698 [nucl-ex]} \BibitemShut
	{NoStop}%
%
	\bibitem [{\citenamefont {Fu}\ and\ \citenamefont
		{Pawlowski}(2015)}]{Fu:2015naa}%
	\BibitemOpen
	\bibfield  {author} {\bibinfo {author} {\bibfnamefont {W.-j.}\ \bibnamefont
			{Fu}}\ and\ \bibinfo {author} {\bibfnamefont {J.~M.}\ \bibnamefont
			{Pawlowski}},\ }\href {\doibase 10.1103/PhysRevD.92.116006} {\bibfield
		{journal} {\bibinfo  {journal} {Phys. Rev. D}\ }\textbf {\bibinfo {volume}
			{92}},\ \bibinfo {pages} {116006} (\bibinfo {year} {2015})},\ \Eprint
	{http://arxiv.org/abs/1508.06504} {arXiv:1508.06504 [hep-ph]} \BibitemShut
	{NoStop}%
%
	\bibitem [{\citenamefont {Almasi}\ \emph
		{et~al.}(2017{\natexlab{b}})\citenamefont {Almasi}, \citenamefont
		{Pisarski},\ and\ \citenamefont {Skokov}}]{Almasi:2016zqf}%
	\BibitemOpen
	\bibfield  {author} {\bibinfo {author} {\bibfnamefont {G.}~\bibnamefont
			{Almasi}}, \bibinfo {author} {\bibfnamefont {R.}~\bibnamefont {Pisarski}}, \
		and\ \bibinfo {author} {\bibfnamefont {V.}~\bibnamefont {Skokov}},\ }\href
	{\doibase 10.1103/PhysRevD.95.056015} {\bibfield  {journal} {\bibinfo
			{journal} {Phys. Rev. D}\ }\textbf {\bibinfo {volume} {95}},\ \bibinfo
		{pages} {056015} (\bibinfo {year} {2017}{\natexlab{b}})},\ \Eprint
	{http://arxiv.org/abs/1612.04416} {arXiv:1612.04416 [hep-ph]} \BibitemShut
	{NoStop}%
%
	\bibitem [{\citenamefont {Skokov}\ \emph {et~al.}(2013)\citenamefont {Skokov},
		\citenamefont {Friman},\ and\ \citenamefont {Redlich}}]{Skokov:2012ds}%
	\BibitemOpen
	\bibfield  {author} {\bibinfo {author} {\bibfnamefont {V.}~\bibnamefont
			{Skokov}}, \bibinfo {author} {\bibfnamefont {B.}~\bibnamefont {Friman}}, \
		and\ \bibinfo {author} {\bibfnamefont {K.}~\bibnamefont {Redlich}},\ }\href
	{\doibase 10.1103/PhysRevC.88.034911} {\bibfield  {journal} {\bibinfo
			{journal} {Phys. Rev. C}\ }\textbf {\bibinfo {volume} {88}},\ \bibinfo
		{pages} {034911} (\bibinfo {year} {2013})},\ \Eprint
	{http://arxiv.org/abs/1205.4756} {arXiv:1205.4756 [hep-ph]} \BibitemShut
	{NoStop}%
%
%
	\bibitem [{\citenamefont {Braun}\ \emph {et~al.}(2004)\citenamefont {Braun},
		\citenamefont {Leonhardt},\ and\ \citenamefont {Pawlowski}}]{Braun:2003ii}%
	\BibitemOpen
	\bibfield  {author} {\bibinfo {author} {\bibfnamefont {J.}~\bibnamefont
			{Braun}}, \bibinfo {author} {\bibfnamefont {K.}~\bibnamefont {Schwenzer}}, \
		and\ \bibinfo {author} {\bibfnamefont {H.~J.}\ \bibnamefont {Pirner}},\
	}\href {\doibase 10.1103/PhysRevD.70.085016} {\bibfield  {journal}
		{\bibinfo  {journal} {Phys. Rev. D}\ }\textbf {\bibinfo {volume} {70}},\
		\bibinfo {pages} {085016} (\bibinfo {year} {2004})},\ \Eprint
	{http://arxiv.org/abs/0312277} {arXiv:0312277 [hep-ph]} \BibitemShut
	{NoStop}%
%
	\bibitem [{\citenamefont {Braun}\ \emph {et~al.}(2019)\citenamefont {Braun},
		\citenamefont {Leonhardt},\ and\ \citenamefont {Pawlowski}}]{Braun:2018svj}%
	\BibitemOpen
	\bibfield  {author} {\bibinfo {author} {\bibfnamefont {J.}~\bibnamefont
			{Braun}}, \bibinfo {author} {\bibfnamefont {M.}~\bibnamefont {Leonhardt}}, \
		and\ \bibinfo {author} {\bibfnamefont {J.~M.}\ \bibnamefont {Pawlowski}},\
	}\href {\doibase 10.21468/SciPostPhys.6.5.056} {\bibfield  {journal}
		{\bibinfo  {journal} {SciPost Phys.}\ }\textbf {\bibinfo {volume} {6}},\
		\bibinfo {pages} {56} (\bibinfo {year} {2019})},\ \Eprint
	{http://arxiv.org/abs/1806.04432} {arXiv:1806.04432 [hep-ph]} \BibitemShut
	{NoStop}%
%
\end{thebibliography}

%

\end{document}